\documentclass[prd,preprint,amsmath,nofootinbib,superscriptaddress]{revtex4}
\usepackage{graphicx,amssymb}
\usepackage{varioref,exscale,latexsym,amsmath,mathrsfs}
\usepackage{youngtab}
\usepackage{bm}
\usepackage{epsfig}
\allowdisplaybreaks

\begin{document}

\title{Multipole expansion at the level of the action} 
\author{Andreas Ross}  
\email{andreasr@andrew.cmu.edu}
\address{Department of Physics, Carnegie Mellon University, 5000 Forbes Ave., Pittsburgh, PA 15213, USA}

\begin{flushright}
\end{flushright}
\begin{abstract}{
Sources of long wavelength radiation are naturally described by an effective field theory (EFT) which takes the form of a multipole expansion. Its action is given by a derivative expansion where higher order terms are suppressed by powers of the ratio of the size of the source over the wavelength. In order to determine the Wilson coefficients of the EFT, i.e. the multipole moments, one needs the mapping between a linear source term action and the multipole expansion form of the action of the EFT. In this paper we perform the multipole expansion to all orders by Taylor expanding the field in the source term and then decomposing the action into symmetric trace free tensors which form irreducible representations of the rotation group. We work at the level of the action, and we obtain the action to all orders in the multipole expansion and the exact expressions for the multipole moments for a scalar field, electromagnetism and linearized gravity. Our results for the latter two cases are manifestly gauge invariant. We also give expressions for the energy flux and the (gauge dependent) radiation field to all orders in the multipole expansion. The results for linearized gravity are a component of the EFT framework NRGR and will greatly simplify future calculations of gravitational wave observables in the radiation sector of NRGR.}
\end{abstract}
\maketitle

\section{Introduction}
The multipole expansion is a standard tool in physics which students usually encounter first when studying electrostatics \cite{jackson}, where it is employed to derive approximate expressions for the electric field at a distance $R$ from a source of size $a$ provided that $a \ll R$ . In this case it is an expansion in $a / R$ and is employed at the level of the solution of a static equation of motion, where its form can be derived from Taylor expanding the Green's function of the Laplacian. In time dependent situations, the multipole expansion does not correspond to an expansion in $a / R$  any more. For the radiation field far away from the source in the regime $R \gg a, \, \lambda$, all multipoles yield a leading contribution to the far zone radiation field falling off as $1/R$. For the leading $1/R$ piece of the radiation field, the multipole expansion is organized as an expansion in $a/\lambda$ which may be truncated for long wavelengths $\lambda \gg a$. 

One important feature of the multipole expansion is that the multipole moments are organized in irreducible representations of the rotation group $\text{SO}(3)$. This has several important advantages. It makes calculations simpler and more transparent, and it ensures the absence of mixing of multipole moments for example in the energy flux in a linear theory (so that we can truly speak of dipole radiation to all orders, otherwise there would be for example dipole--trace-of-octupole radiation at higher orders). Performing the multipole expansion then includes a decomposition into multipole moments which are in irreducible representations of SO(3), where we use symmetric trace free (STF) tensors.

While the traditional field of application of the multipole expansion has been electromagnetism, it is used in general relativity and in particular in gravitational wave physics extensively \cite{Thorne:1980ru, Damour:1990gj, Blanchet:1998in}. In both cases however, it is commonly applied at the level of the solution of the equations of motion when computing the fields. In this paper, we want to study the multipole expansion from a different angle, where we consider the multipole expansion at the level of the action. This is required in many nonrelativistic effective field theories (EFTs) in order to obtain a uniform power counting \cite{Grinstein:1997gv, Goldberger:2004jt}. 

EFTs have become an indispensable tool of modern theoretical physics to systematically study systems with multiple separate scales, see \cite{EFTreview} for a review. Their construction is based on the physical scales in the problem and the underlying symmetries, which together provide the form of the action. It is important to have a uniform power counting for all of the ingredients of the theory, i.e. at the level of the action, so that one can systematically calculate to a given order without having to guess which terms or Feynman diagrams need to be included. The coefficients in front of the terms (or operators in quantum language) in the action are called Wilson coefficients and need to be determined from matching. In weakly coupled theories where the underlying ``full theory'' is known the matching can be performed analytically. Examples for such EFTs include NRQED, NRQCD \cite{NREFT} and NRGR \cite{Goldberger:2004jt} which were constructed to describe nonrelativistic bound states. 

In these theories, the size of the bound state, i.e. the size of the radiation source, is much smaller than the wavelength of emitted radiation. In order to disentangle the different scales and to achieve a uniform power counting, it is necessary to introduce separate modes of the fields which describe physical effects on different scales and which have different kinematic properties. Potential modes for example yield the leading binding dynamics of the source while radiation (or ultrasoft) modes describe physical on-shell radiation. In order to have a uniform power counting of the action written in terms of the different modes, it becomes necessary to Taylor expand the radiation modes describing the physics at the longest distances in the action around a point, which lies within the source \cite{Grinstein:1997gv, Goldberger:2004jt}. This is the origin of the multipole expansion in these nonrelativistic EFTs, where it is required at the level of the action and plays an important role in the matching to an effective radiation theory.

The EFT framework NRGR \cite{Goldberger:2004jt} (see \cite{Goldberger:2007hy} for a pedagogical review and \cite{Goldberger:2006bd} for a brief essay) for nonrelativistic gravitational bound states such as compact binary systems has become an important systematic tool to perform post-Newtonian (PN) computations for gravitational wave templates which are crucial for direct detection experiments such as LIGO/VIRGO \cite{Abramovici:1992ah, Giazotto:1988gw} or eLISA \cite{AmaroSeoane:2012km}. It is an alternative to the more traditional methods to perform PN calculations, see \cite{Blanchet:2002av} for a review.

The EFT construction of NRGR to describe compact binary systems is based on the hierarchy of length scales during the inspiral phase, where the size of the compact objects is much smaller than their orbital separation which in turn is much smaller than the wavelength of the emitted gravitational waves. The first effective description used to describe the physics at distances between the finite size scale and the orbital scale is constructed by replacing the compact objects with point particle worldlines, where their finite size effects can be included with higher order terms in the point particle action. The second stage of the EFT is constructed for scales between the orbital and the wavelength scales, where the potential modes which are responsible for physical effects at the orbital scale are removed from the theory. The resulting theory has a conservative part and a radiation sector in terms of the radiation modes describing gravitational waves.
 
The radiation sector of NRGR has been further developed and explored beyond the leading order in \cite{Goldberger:2009qd}, where the general form of the action for the radiation sector was constructed. Its form is determined by the underlying symmetries, reparameterization invariance and diffeomorphism invariance, and is applicable to arbitrary gravitational wave sources in the long wavelength approximation. This action of the effective long wavelength radiation theory is in the form of a multipole expansion and is a derivative expansion where higher order terms are suppressed by powers of the ratio of the size of the source over the wavelength. The Wilson coefficients of the action, the multipole moments, are not determined by the symmetries and need to be fixed through a matching calculation. 

This matching onto the effective radiation theory is performed perturbatively using Feynman diagrams. In particular, we calculate in the ``full theory'', i.e. in the effective theory which still has potential modes as degrees of freedom, Feynman diagrams with one external radiation mode and arbitrarily many potential modes\footnote{Which diagrams need to be included to a given order is fixed by the power counting of the EFT \cite{Goldberger:2004jt}.} being exchanged between the two point particles. The result of these Feynman diagrams can be written as an effective linear source action $S = -1/(2 m_{Pl}) \int d^4x T^{\mu \nu} \bar h_{\mu \nu}$.  Here, $T^{\mu \nu}$ is the stress-energy pseudo-tensor which includes for example the gravitational binding energy of the system, and we define it in terms of the Feynman diagrams. This action is certainly not in multipole expansion form with a derivative expansion. In fact, to obtain a uniform power counting, it is required that the radiation field in this action is Taylor expanded. This Taylor expansion around a single point forces the resulting radiation theory to be formulated in terms of just one worldline coupled to gravity. So in order to match onto the effective radiation theory, we need to first Taylor expand this action, write it in terms of manifestly gauge invariant operators and finally express the coefficients in irreducible representations of SO(3). This then gives the multipole moments in terms of moments of $T^{\mu \nu}$, which in turn are computed using Feynman diagrams.

In \cite{Goldberger:2009qd} this matching was first done to NLO (for the quadrupole) and resulted in reproducing the energy flux to 1PN. In doing this, the decomposition of the coefficients of the Taylor expansion of the radiation fields into irreducible multipole moments needed to be performed, which at that order was a straightforward exercise. At higher orders \cite{Porto:2010zg}, this decomposition becomes increasingly cumbersome, and the purpose of this paper is to solve it to all orders. Our results will streamline the matching needed in higher order calculations of gravitational wave observables\footnote{NRGR and NRGR-inspired worldline EFTs have produced many interesting results in gravitational wave physics and beyond, see \cite{NRGRpheno}.} in the effective field theory framework NRGR.

In this paper we perform the multipole expansion at the level of the action first for the simple case of a scalar field, then we consider electromagnetism and finally we study general relativity with a linearized source term. While our results for gravity are our main result, we demonstrate our methods used in multipole expanding the action first in the simpler cases. For all three theories studied, we multipole expand the action to all orders and give the exact multipole moments. We also compute the energy flux or radiation power and the radiation field far away from the source in terms of the multipole moments.

The multipole expansion at the level of the solution of the equations of motion has been performed for the same three cases we consider here by Damour and Iyer in \cite{Damour:1990gj}. Their results are given for general sources, but taking the long wavelength limit of their expressions, we find complete agreement with our results for the multipole moments.

{\bf Notation: } We work with a mostly negative metric signature $(+,-,-,-)$. Greek letters denote Lorentz indices ranging from $0 \dots 3$ and lower-case latin indices denote spatial indices from $1 \dots 3$. Upper-case indices denote a set of spatial indices where we use the multi-index notation introduced by Blanchet and Damour \cite{Blanchet:1985sp}. For example, $x^N = x^{k_1} \dots x^{k_n}$ or $x^{ijN-2} = x^{i} x^{j} x^{k_1} \dots x^{k_{n-2}}$. Parenthesis around indices denote symmetrization, for example  $A^{(i} B^j C^{k)} = \frac{1}{3!} \left(A^i B^j C^k + A^i B^k C^j + A^j B^i C^k + A^j B^k C^i + A^k B^i C^j + A^k B^j C^i\right)$. We use natural units $c = \hbar = 1$.

\section{Scalar Field}

\subsection{Multipole expansion of the action}
We first study a scalar field $\phi$ coupled linearly to a source $J$. Its action reads
\begin{align}
S  = \int d^4x \left(\frac{1}{2} \partial_\mu \phi \partial^\mu \phi + J \phi \right)
\end{align}
and its equation of motion is
\begin{equation}
  \Box \phi = J \, .
\end{equation}
Now we consider the configuration where the spatial variation of the field outside the source is much larger than the size of the source $J$, i.e. if spatial derivatives scale as $\partial_i \phi \sim \frac{1}{\lambda} \phi$ and if the size of the source is $a$, we are considering the situation $a \ll \lambda$. For radiation this means we are working in a long wavelength approximation.
In this case we can Taylor expand the field $\phi$ in the source term in the action $S_{\text{source}}$ around a point within the source, and we choose our coordinates such that the point we expand around is the origin $\mathbf x = 0$. 
We plug the Taylor expansion
\begin{equation}
 \phi(t, \mathbf x) = \sum_{n=0}^{\infty} \frac{1}{n !} \, \mathbf x^{k_1} \dots \mathbf x^{k_n} \left(\partial_{k_1} \dots \partial_{k_n} \phi\right)(t, 0) =  \sum_{n=0}^{\infty} \frac{1}{n !} \,  x^N \left(\partial_N \phi\right)(t, 0)
\end{equation}
into the source term in the action 
\begin{equation}
 S_{\text{source}} =   \int d^4x \hspace*{1pt} J \phi = \int dt \int d^3\mathbf x \hspace*{1pt} J(t, \mathbf x) \sum_{n=0}^{\infty} \frac{1}{n !} \,  x^N \left(\partial_N \phi\right)(t,0) = \int dt \sum_{n=0}^{\infty} \frac{1}{n !} \, M^N \partial_N \phi
\end{equation}
where we defined the moments $M^N(t) = \int d^3\mathbf x J(t, \mathbf x) x^N$. Note that all $M^N$ are already symmetric in their indices $k_1 \dots k_n$. In order to bring the source action into the form of a multipole expansion we need to decompose the moments $M^N$ in irreducible representations of the rotation group SO(3) for which we use symmetric trace free (STF) tensors.

The tensors $M^N$ can be expressed in terms of STF tensors starting from the formula for arbitrary symmetric tensors $S^N$ \cite{PiraniSTF, Thorne:1980ru}
\begin{equation}
 S^N = S^N_{\text{STF}} + \sum_{p=1}^{\left[\frac{n}{2}\right]} \frac{(-1)^{p+1} n! (2n - 2p -1)!!}{(n-2p)!(2n-1)!!(2p)!!} \, \delta^{(k_1k_2} \dots \delta^{k_{2p-1}k_{2p}} \hspace*{1pt} S^{k_{2p+1}\dots k_n) a_1 a_1 \dots a_p a_p} \label{MNSTForig}
\end{equation}
where $\left[\frac{n}{2}\right]$ denotes the largest integer $\le n/2$. On the RHS of Eq. (\ref{MNSTForig}) the tensors of lower rank, $S^{k_{2p+1}\dots k_n a_1 a_1 \dots a_p a_p}$, still are not trace free in their free indices $k_{2p+1}\dots k_n $. Therefore, we will use Eq. (\ref{MNSTForig}) recursively to make all tensors trace free, where we mean that all free indices in $\left\{k_1, \dots ,k_n \right\}$ which are not on a $\delta^{k_i k_j}$ should become trace free.
This results in 
\begin{equation}
 S^N = \sum_{p=0}^{\left[\frac{n}{2}\right]} c_p^{(n)} \delta^{(k_1k_2} \dots \delta^{k_{2p-1}k_{2p}} S^{k_{2p+1}\dots k_n) a_1 a_1 \dots a_p a_p}_{\text{STF}} \label{MNSTFnew}
\end{equation}
where the STF prescription only applies to the uncontracted indices $k_{2p+1}\dots k_n $ and where the coefficients are
\begin{equation}
 c_p^{(n)} = \frac{n! (2 n - 4 p + 1)!!}{(2p)!! (n-2p)! (2n-2p+1)!!}. \label{eq_c}
\end{equation}
With this the source action becomes
\begin{align}
 S_{\text{source}} & = \int dt \sum_{n=0}^{\infty} \sum_{p=0}^{\left[\frac{n}{2}\right]} \frac{c_p^{(n)}}{n !} \int d^3 \mathbf x \, J \, r^{2p} x^{N-2P}_{\text{STF}} \left(\nabla^2\right)^p \partial_{N-2P} \phi \notag \\
    & = \int dt \sum_{\ell=0}^{\infty} \frac{1}{\ell ! } \sum_{j=0}^{\infty} \frac{(2\ell + 1)!!}{(2j)!! \,  (2 \ell + 2 j + 1)!!} \int d^3 \mathbf x \, J \, r^{2j} x^{L}_{\text{STF}} \left(\nabla^2\right)^j \partial_{L} \phi 
\end{align}
where we define $r = |\mathbf x|$. Now we may use the equation of motion outside the source $\Box \phi = 0$ to convert the contracted spatial derivatives to time derivatives, and in turn integrate by parts to let them act on the moments rather than on the field. This yields the source action in multipole expanded form
\begin{equation}
 S_{\text{source}} =  \int dt \sum_{\ell=0}^{\infty} \frac{1}{\ell!} \, \mathcal I^L \partial_L \phi \label{eq_MPactionphi}
\end{equation}
with the multipole moments given by
\begin{align}
 \mathcal I^L & = \sum_{p=0}^{\infty} \frac{(2\ell + 1)!!}{(2p)!! \,  (2 \ell + 2 p + 1)!!} \int d^3 \mathbf x \, \partial_t^{2p} J \hspace*{1pt} r^{2p} x^{L}_{\text{STF}} \, .
\end{align}
The normalization of the multipole moments $\mathcal I_L$ is chosen such that for $p=0$ their expression reads
\begin{equation}
 \mathcal I^L_0 = \int d^3 \mathbf x \, J \, x^L_{\text{STF}} \,,
\end{equation}
where we note that the static case of a time independent source is given by only the $p=0$ components of the multipoles.

\subsection{Energy flux for scalar radiation} \label{sec_fluxPhi}
We compute the power or energy flux emitted in scalar waves to all orders in the multipole expansion from Eq. (\ref{eq_MPactionphi}) using the method employed in \cite{Goldberger:2007hy, Goldberger:2009qd}.
For the amputated amplitude of $\phi$ emission from the $\ell$th multipole moment we have
\begin{equation}
 i \mathcal A^{(\ell)} = i \frac{(-i)^{\ell}}{\ell !} \mathcal I^L k^L
\end{equation}
where $k^\mu = (|\mathbf k|, \mathbf k)$ is the outgoing momentum of the emitted on-shell scalar $\phi$.
We then compute the energy flux as in \cite{Goldberger:2007hy, Goldberger:2009qd}
\begin{align}
 \dot E & = \frac{1}{2T} \int \frac{d^3 \mathbf k}{(2 \pi)^3} |\mathcal A|^2 =  \frac{1}{2T} \int \frac{d \Omega_{\mathbf k} \hspace*{1pt} dk \ k^2}{(2 \pi)^3} \sum_{\ell=0}^{\infty} \sum_{\tilde \ell=0}^{\infty} \frac{1}{\ell! \hspace*{1pt} \tilde \ell!} k^{\ell + \tilde \ell} \, \mathcal I^L(k) {\mathcal I^{\tilde L}}^*(k) \, n_{\mathbf k}^L n_{\mathbf k}^{\tilde L} \label{eq_power13}
 \end{align}
where we defined $\mathbf n_{\mathbf k} = \mathbf k / |\mathbf k|$. The angular integral is performed with
\begin{equation}
 \int d\Omega \, n^P = \begin{cases} 0 & \text{if } p \text{ odd} \\ \frac{4 \pi }{(p + 1)} \, \delta^{(k_1 k_2} \dots \delta^{k_{p-1} k_p)} & \text{if } p \text{ even} \end{cases} \label{eq_AngInt}
\end{equation}
which results in the condition $\tilde \ell = \ell$ since the multipole moments are trace free and for $\tilde \ell \neq \ell$ some $\delta$ will contract two indices of the same multipole moment. Moreover,  when using Eq. (\ref{eq_AngInt}), we need to omit the combinations of indices on the $\delta$'s which vanish because the multipole moments are trace free, or in other words, we only need to keep the products of $\delta$'s where each $\delta$ contracts one index on $\mathcal I^L$ with one index on ${\mathcal I^{\tilde L}}^*$. That is accounted for by a factor of $(\ell!)^2 2^\ell / (2\ell)! = \ell!/(2 \ell - 1)!!$ inserted into Eq. (\ref{eq_power13}) and contracting the multipole moments with each other. We find
\begin{align}
 \dot E & =  \frac{1}{4 \pi^2 T} \int_0^\infty dk \sum_{\ell=0}^{\infty} \frac{1}{\ell! (2 \ell + 1)!!} \  k^{2 (\ell+1)} \, \mathcal I^L {\mathcal I^{L}}^* \notag \\
            & =  \sum_{\ell=0}^{\infty} \frac{1}{4 \pi \, \ell! (2 \ell + 1)!!} \left< \left(\frac{d^{\ell+1}}{dt^{\ell+1}} \mathcal I^L\right)^2\right> \label{eq_powerPhi}
\end{align}
where the first line is the expression for the energy flux in momentum space and the second line is the energy flux in coordinate space. In the latter, $<>$ denotes time averaging.

\subsection{Radiation field far from the source} \label{sec_phiField}
At a distance $R \gg \lambda \gg a$ only the $1/R$ component of the field is relevant and we may neglect the components which fall off faster than $1/R$. The field $\phi(t, \mathbf x)$ can be computed from the one-point function, i.e. the $\phi$-emission amplitude with a retarded propagator for the $\phi$ leg. The use of the retarded propagator in diagrammatic field theory calculations can be derived in the in-in formalism \cite{inin}. Its usefulness in the context of NRGR was first emphasized in \cite{Galley:2009px}. For classical calculations of the field (i.e. the one-point function), one can show that the use of retarded propagators is the only modification the in-in formalism implies, even if nonlinearities and multiple couplings to a source are considered. Let us show how to isolate the leading $1/R$ piece of the multipole expanded radiation field.
For that purpose it is convenient to write the multipole expanded action of Eq. (\ref{eq_MPactionphi}) as
\begin{equation}
 S_{\text{source}} =  \int d^4y \sum_{\ell=0}^{\infty} \frac{1}{\ell!} \, \mathcal I^L(y^0) \delta^3(\mathbf y) \frac{\partial}{\partial y^{k_{1}}} \dots \frac{\partial}{\partial y^{k_{\ell}}} \phi (y^0, \mathbf y) \, ,
\end{equation}
and we require the retarded propagator which in d=4 spacetime dimensions reads in coordinate space
\begin{equation}
 \frac{- i \hspace*{1pt} \theta(t_f-t_i) }{4 \pi |\mathbf x_f - \mathbf x_i|} \hspace*{1pt} \delta(t_f - t_i - |\mathbf x_f - \mathbf x_i|) \, .
\end{equation}
Then the field or the one-point function is given by
\begin{align}
 \phi(t, \mathbf x) = i \int d^4y \sum_{\ell=0}^{\infty} \frac{1}{\ell!} \, \mathcal I^L(y^0) \delta^3(\mathbf y) \frac{\partial}{\partial y^{k_{1}}} \dots \frac{\partial}{\partial y^{k_{\ell}}}  \frac{- i \hspace*{1pt} \theta(t-y^0)}{4 \pi |\mathbf x - \mathbf y|} \hspace*{1pt} \delta(t - y^0 - |\mathbf x - \mathbf y|) \label{eq_1pointphi} \, .
\end{align}
Note that the spatial derivatives can either act on the delta function $\delta(t - y^0 - |\mathbf x - \mathbf y|)$ or on the factor $1 / |\mathbf x - \mathbf y|$. For one spatial derivative acting on these two possibilities we have
\begin{align}
\frac{\partial}{\partial y^k} \frac{1}{|\mathbf x - \mathbf y|} & = \frac{(\mathbf x - \mathbf y)^k}{|\mathbf x - \mathbf y|^3} \label{eq_der1}\\
\frac{\partial}{\partial y^k} \delta(x^0-y^0 - |\mathbf x - \mathbf y|) & = \frac{\partial ( x^0 - |\mathbf x - \mathbf y|)}{\partial y^k} \frac{\partial}{\partial(x^0 - |\mathbf x - \mathbf y|)} \delta(x^0-y^0 - |\mathbf x - \mathbf y|) \notag \\
 & = \frac{(\mathbf x -\mathbf y)^k }{|\mathbf x -\mathbf y|} \left(- \frac{\partial}{\partial  y^0} \delta(x^0-y^0 - |\mathbf x - \mathbf y|) \right) \label{eq_der2} \, .
\end{align}
We see that if we integrated these expressions over $\mathbf y$ using the $\delta^3(\mathbf y)$ in the expression of the field in Eq. (\ref{eq_1pointphi}), only the second one where the spatial derivative acts on the delta function would give a term proportional to $1/|\mathbf x|$.
When further derivatives act, they can also act on the vectors $(\mathbf x -\mathbf y)^k$ in Eqs. (\ref{eq_der1}, \ref{eq_der2}). These however will give traces which vanish since the multipole moments are trace free. After integrating over $\mathbf y$, all derivatives acting on powers of $|\mathbf x - \mathbf y|$ will yield terms which fall off faster than $1/|\mathbf x|$. Thus, all spatial derivatives $\nabla_{ \mathbf y}$ have to act on the delta function, where they are effectively converted\footnote{This corresponds to computing the amputated one $\phi$ emission amplitude on-shell in momentum space, since this corresponds to setting $\mathbf k = k^0 \mathbf n$.} to $- \mathbf n \frac{\partial}{\partial y^0}$ where $\mathbf n = \mathbf x / |\mathbf x|$ .
If we then employ this in the expression for the field we find for the radiation field to all orders in the multipole expansion
\begin{align}
 \phi(t, \mathbf x) = \frac{1}{4 \pi |\mathbf x|}\sum_{\ell=0}^{\infty} \frac{1}{\ell!} \, n^L  \, (\partial_t^{\ell} \mathcal I^L)(t_\text{ret}) \label{eq_1pointphiRad}
\end{align}
where we integrated by parts to let all time derivatives act on the multipole moments and where the moments and their time derivatives are evaluated at retarded time $t_\text{ret} = t - |\mathbf x|$.
We can also use the radiation field of Eq. (\ref{eq_1pointphiRad}) to compute the energy flux using $\dot E = \int d\Omega |\mathbf x|^2 \dot \phi^2$ which confirms our result of Eq. (\ref{eq_powerPhi}).

\section{Electromagnetism}

\subsection{Multipole Expansion of the action}
The action for the electromagnetic field $A_\mu$ coupled to a source $J^\mu$ reads
\begin{align}
S & = \int d^4x \left(- \frac{1}{4} F^{\mu \nu} F_{\mu \nu} - J^\mu A_\mu \right)
\end{align}
where $F_{\mu \nu} = \partial_\mu A_\nu - \partial_\nu A_\mu$ and where the current in the source term is conserved,
\begin{equation}
 \partial_\mu J^\mu = 0 \, .
\end{equation} 
The electric and magnetic fields in terms of the potentials are $\mathbf E = - \nabla A^0 - \dot {\mathbf A}$ and $\mathbf B = \nabla \times \mathbf A$ respectively and obey the usual Maxwell equations which are given outside the source by
\begin{equation}
 \nabla \cdot \mathbf E = 0 \, , \ \ \ \ \ \ \
  \nabla \cdot \mathbf B  = 0 \, , \ \ \ \ \ \ \
   \nabla \times \mathbf E = - \dot {\mathbf B} \, , \ \ \ \ \ \ \
   \nabla \times \mathbf B = \dot {\mathbf E} \, .
\end{equation}
Plugging the  Taylor expansion of the field into the source term of the action $S_{\text{source}}$, it reads
\begin{equation}
 S_{\text{source}} =  - \int dt \int d^3\mathbf x J^\mu(t, \mathbf x) \sum_{n=0}^{\infty} \frac{1}{n !} \,  x^N \partial_N A_\mu \, . \label{eq_Ssem1}
\end{equation}
As the first step, we will express $S_{\text{source}}$ in terms of gauge invariant operators. This means that, aside from the monopole term arising from the $\mu = 0$ and $n = 0$ piece, all other components of Eq. (\ref{eq_Ssem1}) will be given in terms of the electric and magnetic fields and their spatial derivatives. For this purpose, we need to consider the $A^0$ and the $A^i$ components separately. The former are simply written as
\begin{align}
 S_{\text{source}}^{A^0}  & =  - \int dt \left(\int d^3\mathbf x J^0\right) A^0 + \int dt \sum_{n=1}^{\infty} \frac{1}{n !} \left(\int d^3\mathbf x J^0 \,  x^N \right) \partial_{N-1} \left(- \nabla_{k_n} A^0\right) \label{eq_SA0nice}.
\end{align}
and we see that, aside from the first term coupling the total charge of the source to the electrostatic potential, the $A^0$ source action includes couplings to the $A^0$ piece of the electric field $\mathbf E$.

Now the $A^i$ piece of the source action,
\begin{align}
 S_{\text{source}}^{A^i} & = \int dt  \sum_{n=0}^{\infty} \frac{1}{n !} \left( \int d^3\mathbf x J^i x^N \right) \partial_N A^i \, , \label{SsourceAi1a} 
\end{align}
needs to be rewritten such that it provides couplings to the magnetic field as well as the $A^i$ components of the electric field which were missing in the $A^0$ part in Eq. (\ref{eq_SA0nice}).
This is achieved by decomposing the source action for $A^i$ into two parts, each with definite symmetry properties in the spatial indices $\{i, k_1 \dots k_n\}$. The first part is totally symmetric in all indices and the second one is antisymmetrized in one pair of indices $(i, k_1)$ and then symmetrized in $\{k_1 \dots k_n\}$. These symmetrizations are conveniently done using Young symmetrizers \cite{YoungSyms} denoted as Young tableaux. In terms of these, the couplings in Eq. (\ref{SsourceAi1a}) are expressed as
\newcommand{\kOneYoung}{k_1}
\newcommand{\kTwoYoung}{k_2}
\newcommand{\kThreeYoung}{k_3}
\newcommand{\kEllYoung}{k_\ell}
\newcommand{\kEnYoung}{k_n}
\begin{equation}
  \int d^3 \mathbf x \, J^i x^N = \frac{1}{(n+1)!} \  {\Yvcentermath1 \young(i\kOneYoung\dots\kEnYoung)} \ + \frac{n}{(n+1)!} \Bigg({\Yvcentermath1 \young(i\kTwoYoung\dots\kEnYoung,\kOneYoung)} +  \text{k-perms}\Bigg) \label{eq_Young1} 
\end{equation}
where 
\begin{align}
 {\Yvcentermath1 \young(i\kOneYoung\dots\kEnYoung)} & = n!  \int d^3 \mathbf x \left[ J^i x^N + \left(J^{k_1} x^{i k_2 \dots k_n} + \text{k-perms}\right) \right] \, , \label{eq_Young1a}\\
 \Bigg({\Yvcentermath1 \young(i\kTwoYoung\dots\kEnYoung,\kOneYoung)} +  \text{k-perms}\Bigg) & = n!  \int d^3 \mathbf x \, J^i x^N - (n-1)!  \int d^3 \mathbf x \left(J^{k_1} x^{i k_2 \dots k_n} + \text{k-perms}\right) \, , \label{eq_Young1b}
\end{align}
and where ``$\! {} + \text{k-perms}$'' means that all other combinations of the indices $\{k_1 \dots k_n\}$ which differ from the original are to be added. For example, it means that in Eq. (\ref{eq_Young1}) there are $n$ Young diagrams to be added, each with a different index $k_j$ in the box in the second row. 
With this decomposition the action becomes
\begin{align}
 S_{source}^{A^i} & = \int dt  \sum_{n=0}^{\infty} \frac{1}{n !} \left(\frac{1}{(n+1)!} \ {\Yvcentermath1 \young(i\kOneYoung\dots\kEnYoung)} \ + \frac{n}{(n+1)!} \Bigg({\Yvcentermath1 \young(i\kTwoYoung\dots\kEnYoung,\kOneYoung)} +  \text{k-perms}\Bigg) \right) \partial_N A^i.
\end{align}
For the totally symmetric piece we now use the conservation of $J^\mu$ from which follows that
\begin{align}
 \int \! d^3 \mathbf x \dot J^{0} x^{iN} = - \! \int \! d^3 \mathbf x \partial_a J^{a} x^{iN} \! = \! \int \! d^3 \mathbf x \left[J^{i} x^N \! + \! \left(J^{k_1} x^{i k_2 \dots k_n} + \text{k-perms}\right) \right] \! = \frac{1}{n!} \!  \ {\Yvcentermath1 \young(i\kOneYoung\dots\kEnYoung)} \label{eq_SymYoung1}
\end{align}
and the part with the antisymmetrization can be written as
\begin{align}
 \Bigg({\Yvcentermath1 \young(i\kTwoYoung\dots\kEnYoung,\kOneYoung)} +  \text{k-perms}\Bigg) & = (n-1)!  \int d^3 \mathbf x \left(  \epsilon^{ik_1c} \epsilon^{abc} J^{a} x^{b k_2 \dots k_n} + \text{k-perms} \right)\, \label{eq_AntiSymYoungEps} .
 \end{align}
Using these expressions in Eqs. (\ref{eq_SymYoung1}) and (\ref{eq_AntiSymYoungEps}), the $A^i$ part of the source action gives
\begin{align}
 S_{\text{source}}^{A^i } & = \! \int \! \hspace*{-1pt} dt \hspace*{-1pt} \sum_{n=1}^{\infty} \frac{1}{n !} \int \! d^3 \mathbf x J^0 x^{N} \partial_{N-1} \!  \left(\! - \dot A^{k_n} \! \right) \!  - \!  \int \! \hspace*{-1pt} dt \hspace*{-1pt} \sum_{n=1}^{\infty}  \frac{n}{(n+1)!} \! \int \!  d^3 \mathbf x  \left(\mathbf J \times  \mathbf x\right)^{k_n} \hspace*{-1pt} x^{N-1} \partial_{N-1} B^{k_n}  \label{eq_SAiSym}
\end{align}
where we integrated by parts to let the time derivative act on the field $A^i$. 
Now we can combine this the source action for $A^0$ adding Eqs. (\ref{eq_SA0nice}) and (\ref{eq_SAiSym}) and obtain
\begin{align}
 S_{\text{source}} & = - \int dt \left(\int d^3\mathbf x J^0\right) A^0 + \int dt  \sum_{n=1}^{\infty} \frac{1}{n !} \int d^3 \mathbf x J^0 x^{N} \partial_{N-1} E^{k_n} \notag \\
  & -  \int dt  \sum_{n=1}^{\infty}  \frac{n}{(n+1)!} \int d^3 \mathbf x \left(\mathbf J \times  \mathbf x\right)^{k_n} x^{N-1} \partial_{N-1}  B^{k_n}  \label{eq_SsourceSym} \, ,
\end{align}
which is in the desired form. Clearly, the source action is now in a manifestly gauge invariant form, however the coefficients of the operators are not yet in irreducible representations of SO(3), i.e. they are not yet in STF form. Bringing the action in a form with coefficients in STF for is performed by first writing all coefficients as symmetric tensors and secondly taking out the traces using the formula of Eq. (\ref{MNSTFnew}). 

The first term in Eq. (\ref{eq_SsourceSym}) is trivially in STF form, whereas the coefficients of the other terms in the first line of Eq. (\ref{eq_SsourceSym}) are already symmetric in all indices $\{k_1 \dots k_n\}$. Thus, we only need to bring the terms in the second line of Eq. (\ref{eq_SsourceSym}),
\begin{align}
 S_{\text{source}}^{(2)} & \equiv - \int dt  \sum_{n=1}^{\infty}  \frac{n}{(n+1)!} \int d^3 \mathbf x \left(\mathbf J \times  \mathbf x\right)^{k_n} x^{N-1} \partial_{N-1}  B^{k_n}  \label{eq_SB} \, ,
\end{align}
in the form of symmetric tensors.  For this purpose we rewrite the decomposition in terms of Young tableaux of Eq. (\ref{eq_Young1}) in the convenient form
\begin{equation}
 K^i x^{N-1} = \left[ K^i x^{N-1}\right]^\text{S} + \frac{1}{n} \left(\epsilon^{i k_1 c} \epsilon^{abc} K^a x^{b k_2 \dots k_{n-1}} + \text{k-perms} \right) \label{eq_symmmm}
\end{equation}
where we made use of Eq. (\ref{eq_AntiSymYoungEps}).
Using this in the action gives
\begin{align}
 S_{\text{source}}^{(2)} & = - \int dt  \sum_{n=1}^{\infty}  \frac{n}{(n+1)!} \int d^3 \mathbf x \left[\left(\mathbf J \times  \mathbf x\right)^{k_n} x^{N-1}\right]^{\text{S}} \partial_{N-1}  B^{k_n}  \notag \\
  & -  \int dt  \sum_{n=2}^{\infty}  \frac{n}{(n+1)!} \, \frac{n-1}{n} \int d^3 \mathbf x \left(\left(\mathbf J \times  \mathbf x\right)\times \mathbf x\right)^{k_{n-1}} x^{N-2} \partial_{N-2} \left(- \nabla \times \mathbf B\right)^{k_{n-1}}  \notag \\
& = - \int dt  \sum_{n=1}^{\infty}  \frac{n}{(n+1)!} \int d^3 \mathbf x \left[\left(\mathbf J \times  \mathbf x\right)^{k_n} x^{N-1}\right]^{\text{S}} \partial_{N-1}  B^{k_n}  \notag \\
  & -  \int dt  \sum_{n=1}^{\infty}  \frac{n}{(n+2)!}  \left[ \int d^3 \mathbf x \ \partial_t \mathbf J \cdot \mathbf x \, x^{N} \right]^\text{S} \partial_{N-1} E^{k_{n}} \notag \\
    & + \int dt  \sum_{n=1}^{\infty}  \frac{n}{(n+2)!}  \left[ \int d^3 \mathbf x \ \partial_t J^{k_{n}} r^2 x^{N-1} \right] \partial_{N-1} E^{k_{n}} \label{eq_S2partly} \, ,
\end{align}
where we have used the Maxwell equations outside the source to replace the curl of $\mathbf B$, and we integrated by parts to let the resulting time derivative act on the moments rather than on the fields. The last term in Eq. (\ref{eq_S2partly}) is still not symmetric, and in order to bring $S_{\text{source}}^{(2)}$ into a form expressed only in terms of symmetric tensors, the symmetrization procedure of Eq. (\ref{eq_symmmm}) has to be applied infinitely many times where we use the Maxwell equations in vacuum to replace any curls of $\mathbf E$ or $\mathbf B$ and integrate all time derivatives by part to act on the moments. This is straightforward since the resulting structures are reoccurring cyclically, and can we derive the expression for $S_{\text{source}}^{(2)}$ in terms of symmetric tensors. With this decomposition, the total source action in terms of symmetric tensors is
\begin{align}
 S_{\text{source}}  & = - \int dt \left(\int d^3\mathbf x J^0\right) A^0 + \int dt  \sum_{n=1}^{\infty} \frac{1}{n !} \left[\int d^3 \mathbf x J^0 x^{N} \right]^S \partial_{N-1} E^{k_n} \notag \\
  & -  \int dt  \sum_{n=1}^{\infty}  \sum_{q=0}^{\infty} \frac{n}{(n+2q+2)!}  \left[ \int d^3 \mathbf x \ \partial_t^{2q+1} \mathbf J \cdot \mathbf x  x^{N} r^{2q} \right]^{\text{S}} \partial_{N-1} E^{k_{n}}  \notag \\
  & +  \int dt  \sum_{n=1}^{\infty} \sum_{q=0}^{\infty} \frac{n}{(n+2q+2)!}  \left[ \int d^3 \mathbf x \ \partial_t^{2q+1} J^{k_{n}} x^{N-1} r^{2q+2} \right]^{\text{S}} \partial_{N-1} E^{k_{n}} \notag \\
  & - \int dt  \sum_{n=1}^{\infty} \sum_{q=0}^{\infty}  \frac{n}{(n+2q+1)!} \left[\int d^3 \mathbf x \, \partial_t^{2q} \left(\mathbf J \times  \mathbf x\right)^{k_n} x^{N-1} r^{2q}\right]^{\text{S}} \partial_{N-1}  B^{k_n} \, . \label{eq_SsourceTotSym}
\end{align}

Now the final step is to use Eq. (\ref{MNSTFnew}) to write the symmetric tensors in terms of STF tensors. The first term in Eq. (\ref{eq_SsourceTotSym}) coupling the static potential to the total charge,
\begin{equation}
  S_{\text{source}}^Q \equiv - \int dt \left(\int d^3\mathbf x J^0\right) A^0 \, , \label{eq_Q}
\end{equation}
is trivially symmetric and trace free. For the other parts, we can essentially repeat the procedure of the scalar case. However, we need to omit the traces which are zero on-shell, i.e. which give terms in the action proportional to $\nabla \cdot \mathbf E = 0$ or $\nabla \cdot \mathbf B = 0$. Moreover, we need to account for different structures of the moments which can arise from traces of the moments. Let us start with the non-trivial part of line 1 of Eq. (\ref{eq_SsourceTotSym}),
\begin{align}
 S_{\text{source}}^{E_1} & \equiv  \int dt  \sum_{n=1}^{\infty} \frac{1}{n !} \left[\int d^3 \mathbf x J^0 x^{N}\right]^S \partial_{N-1} E^{k_n}  \, .
\end{align}
When using Eq. (\ref{MNSTFnew}) for the trace free decomposition of we need to count how many of the symmetrizations  of indices in Eq. (\ref{MNSTFnew}) give terms with $\nabla \cdot \mathbf E$. For a tensor with $n-2p$ free indices on $S_{\text{STF}}$ and $2p$ indices on $\delta$'s on the RHS of Eq. (\ref{MNSTFnew})  there are ${n \choose 2p}$ possible combinations of indices which are on the $\delta$'s (not counting their possible orders). Out of these possibilities, there are ${n-1 \choose p-1}$ possibilities of indices on the $\delta$'s which include the index $k_n$ and which is the index on $E^{k_n}$ in the action. Thus we need to subtract these possibilities when we set $\nabla \cdot \mathbf E = 0$. This gives the piece of the action 
\begin{align}
 S_{\text{source}}^{E_1} & =  \int dt  \sum_{n=1}^{\infty} \sum_{p=0}^{\left[\frac{n}{2}\right]}  \frac{c_p^{(n)}}{n !} \frac{{n \choose 2p} - {n-1 \choose 2p-1}}{{n \choose 2p}} \int d^3 \mathbf x J^0 r^{2p} x^{N-2P}_{\text{STF}} \left(\nabla^2\right)^p \partial_{N-2P-1} E^{k_{n-2p}} \notag \\
  & = \int dt \sum_{\ell=1}^{\infty}  \frac{1}{\ell!} \sum_{p=0}^{\infty} \frac{\ell}{\ell + 2 p} \hspace*{1pt} \frac{(2\ell + 1)!!}{(2p)!! \,  (2 \ell + 2 p + 1)!!} \int d^3 \mathbf x \, \partial_t^{2p} J^0 r^{2p} x^{L}_{\text{STF}}  \partial_{L-1} E^{k_{\ell}} \, , \label{eq_SE}
\end{align}
where we used the wave equation $\Box \mathbf E = 0$ to convert $\nabla^2 \mathbf E$ to $\ddot {\mathbf E}$ and then integrated by parts to let the time derivatives act on the moments rather than on the field.

The trace free decomposition and the counting for the piece of the source action in the second line of Eq. (\ref{eq_SsourceTotSym}) works exactly like the one for $ S_{\text{source}}^{E_1}$ because in both cases all free indices of the moments are on $x^N$. We obtain
\begin{align}
 S_{\text{source}}^{E_2} & \equiv -  \int dt  \sum_{n=1}^{\infty}  \sum_{q=0}^{\infty} \frac{n}{(n+2q+2)!}  \left[ \int d^3 \mathbf x \ \partial_t^{2q+1} \mathbf J \cdot \mathbf x  x^{N} r^{2q} \right]^{\text{S}} \partial_{N-1} E^{k_{n}}  \notag \\ 
  & = -  \int dt  \sum_{\ell=1}^{\infty} \sum_{p=0}^{\infty}  \left(\sum_{j=0}^{p}  c_j^{(\ell + 2 j)} \right) \frac{\ell}{(\ell+2p+2)!} \int d^3 \mathbf x \left( \partial_t^{2p+1} \mathbf J \cdot \mathbf x \right) x^{L}_{\text{STF}} r^{2p} \partial_{L-1} E^{k_{\ell}} \, . \label{eq_SB2}
\end{align}

However when we take traces in the STF decompositions of the remaining terms of Eq. (\ref{eq_SsourceTotSym}), we need to count how many times a factor of $r^2$, $\mathbf J \cdot \mathbf x$, or $\left(\mathbf J \times  \mathbf x\right) \cdot \mathbf x = 0$ occurs. The components of the source action in the third line of  Eq. (\ref{eq_SsourceTotSym}) can be expressed as
\begin{align}
 S_{\text{source}}^{E_3} & \equiv  \int dt  \sum_{n=1}^{\infty} \sum_{q=0}^{\infty} \frac{n}{(n+2q+2)!}  \left[ \int d^3 \mathbf x \ \partial_t^{2q+1} J^{k_{n}} x^{N-1} r^{2q+2} \right]^{\text{S}} \partial_{N-1} E^{k_{n}} \notag \\
  & =  \int dt  \sum_{n=1}^{\infty} \sum_{q=0}^{\infty} \sum_{p=0}^{\left[\frac{n}{2}\right]} \frac{n}{(n+2q+2)!} c_p^{(n)}  \frac{n-2p}{n} \notag \\
   & {} \hspace*{30pt}  \left[ \int d^3 \mathbf x \ \partial_t^{2(q+p)+1} \left(\frac{n-2p}{n} J^{k_{n-2p}} x^{N-2P-1} r^{2(q+p)+2} + \frac{2p}{n} \mathbf J \cdot \mathbf x x^{N-2P} r^{2(q+p)}\right) \right]^{\text{STF}} \notag \\
   & {} \hspace*{38pt} \partial_{N-2P-1} E^{k_{n-2p}} \notag \\
  & =  \int dt  \sum_{\ell=1}^{\infty} \sum_{p=0}^{\infty} \left( \sum_{j=0}^{p}    \frac{c_j^{(\ell + 2j)}}{\ell + 2 j} \right) \frac{\ell^2}{(\ell+2p+2)!}  \left[ \int d^3 \mathbf x \ \partial_t^{2p+1} J^{k_{\ell}} x^{L-1} r^{2p+2} \right]^{\text{STF}} \partial_{L-1} E^{k_{\ell}} \notag \\
    & +  \int dt  \sum_{\ell=1}^{\infty} \sum_{p=0}^{\infty} \left(\sum_{j=0}^{p}    \frac{2j \, c_j^{(\ell + 2j)}}{\ell + 2 j} \right) \frac{\ell}{(\ell+2p+2)!}  \int d^3 \mathbf x \left( \partial_t^{2p+1}  \mathbf J \cdot \mathbf x\right) x_{\text{STF}}^{L} r^{2p}  \partial_{L-1} E^{k_{\ell}} \, .\label{eq_SB3}
\end{align}
The counting of terms from the STF decomposition for the fourth line of Eq. (\ref{eq_SsourceTotSym}) works essentially the same way as for the third line, only that now $\left(\mathbf J \times  \mathbf x\right) \cdot \mathbf x = 0$ and the only possible structure to arise from contracted indices is $r^2$. We find
\begin{align}
 S_{\text{source}}^{B_1} & \equiv - \int dt  \sum_{n=1}^{\infty} \sum_{q=0}^{\infty}  \frac{n}{(n+2q+1)!} \left[\int d^3 \mathbf x \, \partial_t^{2q} \left(\mathbf J \times  \mathbf x\right)^{k_n} x^{N-1} r^{2q}\right]^{\text{S}} \partial_{N-1}  B^{k_n} \notag \\ 
  & =  - \! \int \! dt  \sum_{\ell=1}^{\infty} \sum_{p=0}^{\infty} \hspace*{-1pt}  \left( \sum_{j=0}^{p}  \!  \frac{c_j^{(\ell + 2j)}}{\ell + 2 j} \right) \! \frac{\ell^2}{(\ell \hspace*{-1pt} + \hspace*{-1pt} 2p \hspace*{-1pt} + \hspace*{-1pt} 1)!} \hspace*{-1pt}  \left[ \int \hspace*{-1pt} d^3 \mathbf x \ \partial_t^{2p}  \left(\mathbf J \times  \mathbf x\right)^{k_\ell} x^{L-1} r^{2p} \right]^{\text{STF}} \hspace*{-1pt} \partial_{L-1} B^{k_{\ell}} \, . \label{eq_SB1}
\end{align}

Finally, we can add Eqs. (\ref{eq_Q}), (\ref{eq_SE}), (\ref{eq_SB2}), (\ref{eq_SB3}) and (\ref{eq_SB1}) to find our final result for the multipole expanded electromagnetic source action
\begin{align}
 S_{\text{source}} & = - \int dt \left(\int d^3\mathbf x J^0\right) A^0 + \int dt \sum_{\ell = 1}^{\infty} \frac{1}{\ell!} I^L \partial_{L-1} E^{k_\ell} + \int dt \sum_{\ell = 1}^{\infty} \frac{\ell}{(\ell + 1)!} J^L \partial_{L-1} B^{k_\ell}    
\end{align}
with the exact expressions for the electric and magnetic multipole moments 
\begin{align}
 I^L & =  \sum_{p=0}^{\infty} \frac{(2 \ell + 1)!!}{(2 p )!! (2 \ell+2p+1)!!}  \hspace*{1pt} \frac{\ell}{\ell + 2 p} \left[\int d^3 \mathbf x \, \partial_t^{2p} J^0 r^{2p} x^{L}\right]^{\text{STF}}  \notag \\
       & +  \sum_{p=0}^{\infty} \frac{(2 \ell + 1)!!}{(2 p )!! (2 \ell+2p+1)!!}  \hspace*{1pt} \frac{\ell}{(\ell+1)(\ell + 2 p + 2)} \left[ \int d^3 \mathbf x \, r^{2p} \partial_t^{2p+1} \left( J^{k_{\ell}} x^{L-1} r^{2} - \mathbf J \cdot \mathbf x  x^{L} \right) \right]^{\text{STF}} \label{eq_ILem} \\
J^L & = \sum_{p=0}^{\infty}  \frac{(2 \ell + 1)!!}{(2 p )!! (2 \ell+2p+1)!!}  \left[ \int d^3 \mathbf x \ \partial_t^{2p}  \left(\mathbf x \times  \mathbf J\right)^{k_\ell} x^{L-1} r^{2p} \right]^{\text{STF}}  \, . \label{eq_JLem}
\end{align}
We note that we used Eq. (\ref{eq_Sum1}) to eliminate the sums over $j$ present in Eqs. (\ref{eq_SB2} - \ref{eq_SB1}), and we chose a normalization\footnote{Note that our normalization differs from some of the usual conventions used in electrodynamics (see e.g. \cite{jackson}) because we adapt the conventional normalization for multipole moments in gravitational wave physics.} such that the electric moments in the static limit where $p = 0$ are $I^L_0 = \int d^3 \mathbf x  J^0 x^{L}_{\text{STF}}$ and the magnetic moments for $p =0$ are given by $J^L_0 = \left[\int d^3 \mathbf x \left(\mathbf x \times \mathbf J\right)^{k_{\ell}} x^{L-1}\right]^{\text{STF}}$. Due to the conservation of the electromagnetic current $\partial_\mu J^\mu = 0$, our expression for the electric multipole moments in Eq. (\ref{eq_ILem}) is not unique, since we can add or subtract terms which vanish due to current conservation. In particular, current conservation implies
\begin{equation}
 \int d^3 \mathbf x \, \partial_t J^0 r^{2p} x^{L} = \left[ \int d^3 \mathbf x \, r^{2p-2} \left(\ell \hspace*{1pt} J^{k_{\ell}} x^{L-1} r^{2} + 2 p \hspace*{1pt} \mathbf J \cdot \mathbf x \hspace*{1pt} x^{L} \right) \right]^{\text{S}},
\end{equation}
which can be used to write $p > 0$ pieces of $I^L$ in the first line of Eq. (\ref{eq_ILem}) in terms of the form of the components seen in the second line and vice versa.

\subsection{Radiated power}
We calculate the energy flux or radiated power in the same way as above for the scalar field in Sec. \ref{sec_fluxPhi}. In order to compute the total energy flux however, we need to sum the amputated photon emission amplitude squared over all polarizations $h$, and one needs to choose a gauge to have an explicit expression for the polarization sum. Since it is slightly simpler we will employ transverse gauge or Coulomb gauge, where the polarization vectors satisfy $\epsilon^0(\mathbf k, h) = 0$ and $\mathbf k \cdot \epsilon(\mathbf k, h) = 0$ and where the polarization sum is
\begin{equation}
 \sum_h \epsilon^i(\mathbf k, h) \epsilon^{j*}(\mathbf k, h) = \delta^{ij} - \mathbf k^i \mathbf k^j / \mathbf k^2 \, . \label{eq_EMpolSum}
\end{equation}
In this gauge there is no $A^0$ emission, and the amputated emission amplitude is
\begin{equation}
 i \mathcal A_h = i \sum_{\ell = 1}^\infty  \frac{(-i)^\ell}{\ell!} I^L(|\mathbf k|) \, k^{L-1} |\mathbf k| \epsilon^{k_\ell *}(\mathbf k, h) + i \sum_{\ell = 1}^\infty \frac{(-i)^\ell \ell}{(\ell + 1)!}  J^L(|\mathbf k|) \, k^{L-1} k^i \epsilon^{ijk_\ell} \epsilon^{j *}(\mathbf k, h) \, .
\end{equation}
Squaring the amplitude, summing over helicities with Eq. (\ref{eq_EMpolSum}) and performing the angular integration with the help of Eq. (\ref{eq_AngInt}) yields the energy flux
\begin{align}
 \dot E & =  \frac{1}{4 \pi^2 T} \hspace*{-1pt}  \int_0^\infty \! dk \hspace*{-1pt}  \left[\sum_{\ell=1}^{\infty} \frac{\ell + 1}{\ell \, \ell! (2 \ell+ 1)!!} \,  k^{2 (\ell+1)} \hspace*{-1pt} \left| I^L(k)\right|^2 + \sum_{\ell=1}^{\infty}  \frac{\ell}{(\ell + 1)! (2 \ell + 1)!!} \,  k^{2 (\ell+1)} \hspace*{-1pt} \left| J^L(k)\right|^2 \right] \notag \\
            & =  \sum_{\ell=1}^{\infty} \frac{\ell + 1}{4 \pi \ell \, \ell! (2 \ell + 1)!!} \left< \left(\frac{d^{\ell+1}}{dt^{\ell+1}} I^L\right)^2\right>  + \sum_{\ell=1}^{\infty} \frac{\ell}{4 \pi (\ell + 1)! (2 \ell + 1)!!} \left< \left(\frac{d^{\ell+1}}{dt^{\ell+1}} J^L\right)^2\right> \label{eq_powerEM} \, .
\end{align}
It is straightforward to check that if we use the covariant Feynman gauge instead where we need to keep $A^0$, we obtain the same result for the energy flux. This is of course expected because our multipole moments are gauge independent and the energy flux is an observable and cannot depend on the gauge.

\subsection{Electromagnetic radiation field}
In calculating the radiation field we will work in Feynman gauge where the retarded photon propagator is given by
\begin{equation}
 \frac{i \eta_{\mu\nu} \hspace*{1pt} \theta(t_f-t_i) }{4 \pi |\mathbf x_f - \mathbf x_i|} \hspace*{1pt} \delta(t_f - t_i - |\mathbf x_f - \mathbf x_i|) \, .
\end{equation}
Using the one-point function we can then calculate the radiation fields $A^0(t, \mathbf x)$ and $A^i(t, \mathbf x)$ as discussed in more detail for the scalar field in Sec. \ref{sec_phiField}.
We obtain for the multipole expanded radiation field in Feynman gauge
\begin{align}
 A^0(t, \mathbf x) & = \frac{1}{4 \pi |\mathbf x|}\sum_{\ell=1}^{\infty} \frac{1}{\ell!} \, n^L  \, (\partial_t^{\ell}  I^L)(t_\text{ret}) \label{eq_1pointA0Rad} \\
  A^i(t, \mathbf x) & = \frac{1}{4 \pi |\mathbf x|} \left[\sum_{\ell=1}^{\infty} \frac{1}{\ell!} \, n^{L-1}  \, (\partial_t^{\ell}  I^{iL-1})(t_\text{ret}) - \sum_{\ell=1}^{\infty} \frac{\ell}{(\ell+1)!} \, \epsilon^{ijk_\ell} n^{jL-1}  \, (\partial_t^{\ell}  J^{L})(t_\text{ret}) \right] \label{eq_1pointAiRad} \, .
\end{align}

\section{Linearized gravity and NRGR} \label{sec_gr}
For gravity our conventions are $m_{Pl}^2 = 1/(32 \pi G)$ for the normalization of the Planck mass relative to Newton's constant, the Riemann tensor is ${R^\mu}_{\nu \alpha \beta} =
\partial_\alpha \Gamma^{\mu}_{\nu \beta} - \partial_\beta
\Gamma^\mu_{\nu \alpha} + \Gamma^\mu_{\lambda \alpha}
\Gamma^\lambda_{\nu \beta} - \Gamma^\mu_{\lambda \beta}
\Gamma^\lambda_{\nu \alpha}$, the Ricci tensor is $R_{\mu \nu} = {R^\alpha}_{\mu
\alpha \nu}$, and we remind the reader that we work with a mostly negative metric signature.
When we expand the metric around flat Minkowski spacetime, we normalize the gravitational field as $g_{\mu\nu} = \eta_{\mu \nu} + h_{\mu \nu} / m_{Pl}$.

The action we want to study is given by the standard Einstein-Hilbert term plus a linear source term. It reads
\begin{align}
S & = - 2 m_{Pl}^2 \int d^4x \sqrt{- g} R - \frac{1}{2 m_{Pl}} \int d^4x T^{\mu \nu} h_{\mu \nu} \, ,
\end{align}
where $\partial_\mu T^{\mu \nu} = 0$, and the vacuum equations of motion are
\begin{equation}
 R_{\mu \nu} = 0 \, .
\end{equation}

The linear gravitational source $T^{\mu \nu}$ may be either the standard stress-energy tensor of matter or it may be the stress-energy pseudo-tensor which includes some gravitational effects such as the gravitational binding energy of a bound state. The latter is the case in the effective field theory framework NRGR \cite{Goldberger:2004jt}, where different kinematic regions or modes of the gravitational field $h_{\mu \nu}$ are treated separately by writing $h_{\mu \nu} = \bar h_{\mu \nu} + H_{\mu \nu}$. The radiation modes $\bar h_{\mu \nu}$ describe on-shell gravitational waves whereas the potential modes $H_{\mu \nu}$ are responsible for the conservative binding dynamics of a bound state such as a binary system. 

The characteristic length scale for the potential modes is the size of the source $a$, whereas for the radiation modes it is the wavelength $\lambda \gg a$. Therefore if we consider physical effects at distances much larger than the size of the source $a$, the potential modes are not needed as explicit degrees of freedom in the theory. Integrating out the potential modes, i.e. solving for them and replacing them by their solutions in the action, we obtain an effective radiation theory only in terms of the radiation modes. The effective action of this effective radiation theory is expanded in powers of the radiation field $\bar h_{\mu \nu}$,
\begin{equation}
S_{\text{eff}} = S_0 + S_1 + \mathcal O(\bar h^2)
\end{equation}
 and is obtained perturbatively using Feynman diagrams. The action $S_0$ of zeroth order in $\bar h_{\mu \nu}$, obtained from Feynman diagrams with no radiation modes and with any number of potential modes exchanged between the two wordlines describing the two binary constitutents, gives the action describing the conservative dynamics of the system. The first order couplings in $S_1$ are obtained from Feynman diagrams with any number of potential modes being exchanged and one external radiation mode and can be written as a linear source action\footnote{Terms in the effective action with coupling of quadratic and higher order in the radiation field become relevant for higher order calculations, in post-Newtonian applications they start at order $v^5$ or 2.5PN.}. Thus when we use the term linearized gravity in this paper, what we mean is that the source term in the action is only coupling linearly to the radiation field. In other words, we treat the coupling of the effective source $T^{\mu \nu}$ to the long wavelength modes linearly, while the interactions of short-distance modes (which are implicit in $T^{\mu \nu}$) can be included to arbitrary order.

\subsection{Multipole expansion of the action}
As in the previous sections, we Taylor expand the gravitational field $h_{\mu \nu}$ in the source action $S_{\text{source}}$ around a point within the source choosing our coordinates such that the point we expand around is the origin $\mathbf x = 0$. 
The source term in the action then is
\begin{align}
 S_{\text{source}} & = - \frac{1}{2 m_{Pl}} \int d^4x  T^{\mu \nu} h_{\mu \nu} = - \frac{1}{2 m_{Pl}} \int dt  \sum_{n=0}^{\infty} \frac{1}{n !} \,  \int d^3\mathbf x T^{\mu \nu}(t, \mathbf x) x^N \left(\partial_N h_{\mu \nu}\right)(t, 0) \, .
\end{align}
As in the electromagnetic case, our first step will be to express this Taylor expanded source action in terms of manifestly gauge invariant operators. In general relativity, the gauge invariant quantities are the Riemann tensor, the Ricci tensor and the Ricci scalar and their covariant derivatives. Due the vacuum equations of motion $R_{\mu \nu} = R = 0$ however, only structures involving the Riemann tensor  $R_{\mu \nu \rho \sigma}$ yield physical terms which can contribute to observable. Terms with Ricci tensor or scalar can be removed via field redefinitions \cite{Goldberger:2004jt, Blanchet:2005tk}; in EFT jargon such terms are called redundant operators. Now when writing the Riemann tensor in terms of temporal and spatial indices, there are only three distinct combinations of indices: $R_{0i0j}$, $R_{0ijk}$ and $R_{ijkl}$. In linearized gravity however, the purely spatial Riemann tensor is on-shell (i.e. upon use of the vacuum equations of motion)
\begin{align}
R_{ijkl} & = - \epsilon_{ijm} \epsilon_{klp} R_{0m0p} \, , \label{eq_puSpRiemann}
\end{align}
so that we may replace it by $R_{0i0j}$. The couplings to the two remaining components of the Riemann tensor, $R_{0i0j}$ and $R_{0ijk}$, are conveniently organized in terms of the electric and magnetic components of the Riemann tensor\footnote{On-shell, where $R_{\mu \nu}$, the Riemann tensor $R_{\mu \nu \rho \sigma}$ and the Weyl tensor $C_{\mu \nu \rho \sigma}$ used in previous work \cite{Goldberger:2009qd} coincide.}. They are given by
\begin{align}
 E_{ij}& = R_{0i0j}= \frac{1}{2m_{Pl}}\left(\partial_0 \partial_j h_{0i} + \partial_0 \partial_i h_{0j} - \partial_i \partial_j h_{00} - \partial_0^2 h_{ij}\right) \label{eq:WeylEnonrel} \\
 B_{ij} & = \frac{1}{2} \epsilon_{imn} R_{0jmn} = \frac{1}{2 m_{Pl}} \epsilon_{imn} \left(\partial_0 \partial_n h_{jm} + \partial_j \partial_m h_{0n} \right) \, ,
\end{align}
where the last expressions are expanded to linear order in $h_{\mu \nu}$. Note that we gave the expressions for $E_{ij}$ and $B_{ij}$ in a system with four-velocity $v^\mu  = (1,0,0,0)$, which makes sense for our application to the multipole expansion where we Taylor expand around the point $\mathbf x = 0$ which implies the source is described by a worldline $x^\mu = (t, 0)$ whose four-velocity is $v^\mu  = (1,0,0,0)$.

The electric and magnetic components of the Riemann tensor in vacuum have the properties
$E_{ij} = E_{ji}$, $E_{ii} = 0$, $B_{ij} = B_{ji}$ and $B_{ii} = 0$,
i.e. they are symmetric and trace free.
Furthermore, derivatives acting on the electric and magnetic components of the Riemann tensor in vacuum give the equations of motion which are in complete correspondence to Maxwell's equations,
\begin{align}
 \partial_i E_{ij}  =  0, \ \ \ \ \ \ \ \
 \partial_i B_{ij}  = 0, \ \ \ \ \ \ \ \
\epsilon_{imn} \partial_m E_{jn}  = \dot B_{ij}, \ \ \ \ \ \ \ \
 \epsilon_{imn} \partial_m B_{jn}  =  - \dot E_{ij} \, ,
 \end{align}
which imply the wave equations
 \begin{align}
 \Box E_{ij}  = 0 \ \ \ \ \Box B_{ij}  = 0 \, .
\end{align}

Having reviewed these properties of the electric and magnetic components of the Riemann tensor, we set out to write the Taylor expanded action in terms of $E_{ij}$ and $B_{ij}$. Similar to the electromagnetic case where the coupling to the total (conserved) charge could not be expressed as a coupling to $\mathbf E$ or $\mathbf B$, we also expect that the couplings to the leading multipoles in gravity cannot to be expressed in terms of $E_{ij}$ and $B_{ij}$. 

In order to express the Taylor expanded action mostly in terms of $E_{ij}$ and $B_{ij}$, we have to consider the couplings to $h_{00}$, $h_{0i}$ and $h_{ij}$ separately.
First we write the $h_{00}$ part of the source action as
\begin{align}
 S_{\text{source}}^{h_{00}} & = - \frac{1}{2 m_{Pl}} \int dt \left[ \left(\int d^3\mathbf x T^{00}\right)  h_{00} + \left( \int d^3\mathbf x T^{00} x^k \right) \partial_k h_{00} \right] \notag \\
   & + \int dt \sum_{n=2}^{\infty} \frac{1}{n !} \left( \int d^3\mathbf x T^{00}(t, \mathbf x)   x^N \right) \partial_{N-2} \left(\frac{1}{2 m_{Pl}} \left(-\partial_{k_{n-1}} \partial_{k_{n}}h_{00}\right)\right) \, .
\end{align}
where the first two terms are the couplings to the total energy and to the center of mass coordinate (times the total energy) respectively. 
The $h_{0i}$ part reads
\begin{align}
 S_{\text{source}}^{h_{0i}} & = - \frac{1}{m_{Pl}} \int dt \left[ \left(\int d^3\mathbf x T^{0i}\right)  h_{0i} + \left( \int d^3\mathbf x T^{0i} x^k \right) \partial_k h_{0i} \right] \notag \\
   & + \int dt \sum_{n=2}^{\infty} \frac{1}{n !} \left( \int d^3\mathbf x T^{0i}(t, \mathbf x)   x^N \right) \partial_{N-2} \left(- \frac{1}{m_{Pl}} \partial_{k_{n-1}} \partial_{k_{n}}h_{0i}\right) \, .
\end{align}
where the first term is the coupling to the total linear momentum and the second term contains the total angular momentum.
All terms besides the first one need to be decomposed to forms of definite symmetry using Young tableaux, where we may use our results from the E\&M calculation for the vector potential in Eqs. (\ref{eq_Young1} - \ref{eq_Young1b}) with $J^i$ replaced by $T^{0i}$.
For the totally symmetric piece we then use the conservation of $T^{\mu \nu}$ from which follows that
\begin{align}
 \int \! d^3 \mathbf x \dot T^{00} x^{iN} = - \! \int \! d^3 \mathbf x \partial_a T^{0a} x^{iN} \! = \! \int \! d^3 \mathbf x \left[ T^{0i} x^N \! + \! \left(T^{0 k_1} x^{i k_2 \dots k_n} + \text{k-perms}\right) \right] \! = \frac{1}{n!} \!  \ {\Yvcentermath1 \young(i\kOneYoung\dots\kEnYoung)} \label{eq_SymYoung1GR}
\end{align}
and for the part with the antisymmetrization we use
\begin{align}
 \Bigg({\Yvcentermath1 \young(i\kTwoYoung\dots\kEnYoung,\kOneYoung)} +  \text{k-perms}\Bigg) & = (n-1)!  \int d^3 \mathbf x \left(\epsilon^{ik_1c} \epsilon^{abc} T^{0a} x^{b k_2 \dots k_n} + \text{k-perms}  \right) \, \label{eq_AntiSymYoungEpsB} .
 \end{align}
Note that Eqs. (\ref{eq_SymYoung1GR}) and (\ref{eq_AntiSymYoungEpsB}) and also are identical to Eqs. (\ref {eq_SymYoung1}) and (\ref{eq_AntiSymYoungEps}) in the decomposition for the electromagnetic case, with $(J^0,J^a)$ replaced by $(T^{00},T^{0a})$.

With these the $h_{0i}$ part of the action becomes
\begin{align}
 S_{\text{source}}^{h_{0i}} & = - \frac{1}{m_{Pl}} \int dt \left[ \left(\int d^3\mathbf x T^{0i}\right)  h_{0i} + \left\{\frac{1}{2} \int d^3\mathbf x \left(T^{0i} x^k - T^{0k} x^i \right)\right\} \frac{1}{2} \left(\partial_k h_{0i} -  \partial_i h_{0k}\right) \right] \notag \\
   & + \int dt \sum_{n=2}^{\infty} \frac{1}{n !} \left( \int d^3\mathbf x T^{00} x^N \right) \partial_{N-2} \left(\frac{1}{2 m_{Pl}} \left(\partial_0 \partial_{k_{n-1}} h_{0k_n} + \partial_0 \partial_{k_{n}} h_{0k_{n-1}}\right) \right) \notag \\
  & - \int dt \sum_{n=2}^{\infty} \frac{2 n}{(n+1) !} \left( \int d^3\mathbf x \epsilon^{k_n ba}T^{0a} x^{b k_1 \dots k_{n-1}} \right) \partial_{N-2} \left(\frac{1}{2 m_{Pl}} \left(\epsilon_{k_n cd} \partial_{k_{n-1}} \partial_c h_{0d}\right) \right) \, .
\end{align}
For the  $h_{ij}$ part of the source action,
\begin{align}
 S_{\text{source}}^{h_{ij}} & = - \frac{1}{2 m_{Pl}} \int dt \sum_{n=0}^{\infty} \frac{1}{n !} \int d^3\mathbf x T^{ij} x^N \partial_N h_{ij} \, ,
\end{align}
the decomposition in Young tableaux reads
\begin{align}
 \int d^3 \mathbf x \, T^{ij} x^N & = \frac{1}{(n+2)!} \  {\Yvcentermath1 \young(ij\kOneYoung\dots\kEnYoung)} \ + \frac{n+1}{(n+2)!} \Bigg({\Yvcentermath1 \young(ij\kTwoYoung\dots\kEnYoung,\kOneYoung)} +  \text{k-perms}\Bigg) \notag \\
  & + \frac{n-1}{2(n+1)!} \Bigg({\Yvcentermath1 \young(ij\kThreeYoung\dots\kEnYoung,\kOneYoung\kTwoYoung)} +  \text{k-perms}\Bigg) \label{eq_YoungGR} \, . 
\end{align}
The first Young tableau is given by
\begin{align}
& {\Yvcentermath1 \young(ij\kOneYoung\dots\kEnYoung)}  \notag \\ & = 2 \, n!  \int \hspace*{-1pt} d^3 \mathbf x \left[T^{ij} x^N + \left(\big(T^{ik_n} x^{jN-1}  +  T^{jk_n} x^{N-1}\big)+  \text{k-perms}\right) +  \left(T^{k_n k_{n-1}} x^{ijN-2} +  \text{k-perms}\right)\right] \notag \\
  & = n! \int d^3 \mathbf x \,  \ddot T^{00} x^{ijN} \, , \label{eq_YT1}
\end{align}
where we used the conservation of $T^{\mu \nu}$ to derive the last equality, and the second one is
\begin{align}
 & {\Yvcentermath1 \young(ij\kTwoYoung\dots\kEnYoung,\kOneYoung)} +  \text{k-perms} \notag \\ 
& = (n-1)! \int d^3 \mathbf x \Big[ 2 n \hspace*{1pt} T^{ij} x^N +  (n-2) \left(\big(T^{ik_n} x^{jN-1} + T^{jk_n} x^{iN-1}\big) +  \text{k-perms}\right) \notag \\
& {} \hspace*{90pt}  - 4 \left(T^{k_n k_{n-1}} x^{ijN-2} +  \text{k-perms}\right)\Big] \notag \\
 & = (n-1)!  \int d^3 \mathbf x \left[\epsilon^{i k_n c} \epsilon^{abc} \left\{T^{aj} x^{bN-1} + \left( T^{a k_{n-1}}x^{bjN-2} + \text{k-perms}\right)\right\} + (i \leftrightarrow j)\right] + \text{k-perms} \label{eq_YoungGRantisym}  \, .
\end{align}
In the last line of Eq. (\ref{eq_YoungGRantisym}), the first $+ \text{k-perms}$ inside the parenthesis means that each of the indices $k_i$ which are not on the epsilon tensor, i.e. which are inside the round parenthesis, is to be used once as an index on $T^{ak_i}$ inside the round parenthesis. The second $+ \text{k-perms}$ in Eq. (\ref{eq_YoungGRantisym}) means that each $k_i$ is to be used once as an index on the epsilon tensor.
The third Young tableau is
\begin{align}
&  {\Yvcentermath1 \young(ij\kThreeYoung\dots\kEnYoung,\kOneYoung\kTwoYoung)} +  \text{k-perms} \notag \\
& = 2 \, (n-2)! \int d^3 \mathbf x \Big[ n(n-1) T^{ij} x^N -  (n-1) \left(\big(T^{ik_n} x^{jN-1} + T^{jk_n} x^{N-1}\big) +  \text{k-perms}\right) \notag \\
& {} \hspace*{96pt}  + 2 \left(T^{k_n k_{n-1}} x^{ijN-2} +  \text{k-perms}\right)\Big] \notag \\
& = 2 \, (n-2)! \left[\left\{\epsilon^{cik_n}\epsilon^{djk_{n-1}}\epsilon^{cab}\epsilon^{def} \int d^3\mathbf x T^{ae} x^{bfN-2} + (i \leftrightarrow j)\right\} + \text{k-perms}\right] \label{eq_YT3} \, .
\end{align}
When we plug these into the $h_{ij}$ part of the source action we obtain
\begin{align}
 S_{source}^{h_{ij}} & =  \int dt \sum_{n=2}^{\infty} \frac{1}{n!} \int d^3\mathbf x T^{00} x^{N} \partial_{N-2} \left(\frac{1}{2 m_{Pl}} \left(- \partial_0^2 h_{k_{n-1} k_n} \right) \right) \notag \\
  & - \int dt \sum_{n=2}^{\infty} \frac{2n}{(n+1)!} \int d^3\mathbf x \epsilon^{k_nba} T^{0a} x^{bN-1} \partial_{N-2} \left(\frac{1}{2 m_{Pl}} \left(\epsilon_{k_ncd} \partial_0 \partial_d h_{c k_{n-1}} \right) \right) \notag \\
  & + \int dt \sum_{n=2}^{\infty} \frac{n-1}{(n+1)!} \int d^3\mathbf x \left[T^{aa} x^{N} + T^{k_{n-1} k_n} \mathbf x^2 x^{N-2} - 2 T^{k_na} x^{N-1 \hspace*{1pt} a}\right] \partial_{N-2} E_{k_{n-1} k_n}  \label{eq_Sshij}
\end{align}
where we have used
\begin{align}
 \int d^3\mathbf x \dot T^{0a} x^{bN} =  \int d^3\mathbf x \left[T^{ab} x^N + \left(T^{ak_n} x^{bN-1} + \text{k-perms} \right)\right]
\end{align}
to arrive at the expression in the second line of Eq. (\ref{eq_Sshij}), and we used the epsilon tensors from Eq. (\ref{eq_YT3}) to write $\partial_{k_{n-1}} \partial_{k_{n}} h_{ij}$ as the purely spatial Riemann tensor, which in turn is written as the electric components of the Riemann tensor via Eq. (\ref{eq_puSpRiemann}).

Adding all components of the source action, the total becomes
\begin{align}
 S_{\text{source}} & =  S_{\text{source}}^{cons} + S_{\text{source}}^{rad}
\end{align}
with
\begin{align}
 S_{\text{source}}^{cons} & = - \frac{1}{2 m_{Pl}} \int dt \left[ \left(\int d^3\mathbf x T^{00}\right)  h_{00} + 2 \left(\int d^3\mathbf x T^{0i}\right)  h_{0i}  \right] \notag \\
  & - \frac{1}{2 m_{Pl}} \int dt \left[ \left( \int d^3\mathbf x T^{00} x^k \right) \partial_k h_{00} + \left(- \int d^3\mathbf x \left(T^{0i} x^k - T^{0k} x^i \right)\right) \frac{1}{2} \left(\partial_i h_{0k} -  \partial_k h_{0i}\right) \right] \notag \\
  & = - \frac{1}{2 m_{Pl}} \int dt \left[M  h_{00} + 2 \mathbf P^i h_{0i} + M \mathbf X^i \partial_i h_{00} + \mathbf L^i \epsilon_{ijk} \partial_j h_{0k}\right] \\
S_{\text{source}}^{rad} & = \int dt \sum_{n=2}^{\infty} \frac{1}{n!}  \int d^3\mathbf x T^{00} x^{N} \hspace*{1pt}  \partial_{N-2} E_{k_{n-1} k_n} \notag \\
  & + \int dt \sum_{n=2}^{\infty} \frac{n-1}{(n+1)!} \int d^3\mathbf x \left[T^{aa} x^{N} + T^{k_{n-1} k_n} r^2 x^{N-2} - 2 T^{k_na} x^{N-1 \hspace*{1pt} a}\right] \partial_{N-2} E_{k_{n-1} k_n} \notag \\
  & - \int dt \sum_{n=2}^{\infty} \frac{2n}{(n+1)!} \int d^3\mathbf x \epsilon^{k_nba} T^{0a} x^{bN-1} \, \partial_{N-2} B_{k_{n-1} k_n} \, .
\end{align}
We note that $S_{\text{source}}^{cons}$, which is not written in terms of $E_{ij}$ and $B_{ij}$, couples to several quantities which are conserved due to the conservation of $T^{\mu \nu}$: The electric parity mass monopole $M \equiv \int d^3\mathbf x T^{00}$ is the total energy (ADM mass) of the source\footnote{Note the change in notation from \cite{Goldberger:2009qd} where the mass monopole was denoted $m$.}, $\mathbf P^i \equiv \int d^3\mathbf x T^{0i}$ is the total linear momentum (which interestingly does not have a consistent classification as a multipole moment since some of its aspects are characteristic of a monopole such as its coupling without derivative, but it is a vector like a dipole), and the magnetic parity current dipole $L^{ij} \equiv - \int d^3\mathbf x \left(T^{0i} x^j - T^{0j} x^i \right)$ is the total angular momentum (which is related to the usual angular momentum 3-vector via $L^{ij} = \epsilon^{ijk} \mathbf L^k$). The only coupling in $S_{\text{source}}^{cons}$ which is not conserved is the mass dipole, related to the center of mass position $\mathbf X^i \equiv 1/M \int d^3\mathbf x T^{00} x^i$. However, $\partial_\mu T^{\mu \nu} = 0$ implies $M \dot {\mathbf X}^i = \mathbf P^i$ and $\ddot {\mathbf X}^i = 0$. Thus, none of these quantities can radiate since the energy flux would be proportional to the square of the first time derivatives of $M$ and $\mathbf P^i$ or the second time derivatives of the dipoles. Moreover, it is straightforward to show that under coordinate transformations $S_{\text{source}}^{cons}$ only changes by total time derivatives which leave the dynamics unchanged. It usually is convenient to work in the center of mass frame where $\mathbf X^i = \mathbf P^i = 0$. 

The couplings in the second part of the source action $S_{\text{source}}^{rad}$ are all able to emit radiation, and we now set out to decompose $S_{\text{source}}^{rad}$ in terms of STF tensors. As done in the previous two sections, we first write all structures as symmetric tensors. For this purpose we write
\begin{align}
S_{source}^{rad} & = S_0 + S_1^{E, \, 0} + S_1^{B, \, 0} + S_2
\end{align}
with
\begin{align}
S_0 & = \int dt \sum_{n=2}^{\infty} \frac{1}{n!} \int d^3\mathbf x \left\{T^{00} + \frac{n-1}{n+1} T^{aa} \right\} x^{N} \partial_{N-2} E_{k_{n-1} k_n} \\
S_1^{E, \, 0} & = - \int dt \sum_{n=2}^{\infty} \frac{2(n-1)}{(n+1)!} \int d^3\mathbf x \, T^{k_na} x^{a N-1} \, \partial_{N-2} E_{k_{n-1} k_n}  \\
S_1^{B, \, 0} & = - \int dt \sum_{n=2}^{\infty} \frac{2n}{(n+1)!} \, \int d^3\mathbf x \epsilon^{k_nba} T^{0a} x^{bN-1} \, \partial_{N-2} B_{k_{n-1} k_n} \\
S_2 & =  \int dt \sum_{n=2}^{\infty} \frac{n-1}{(n+1)!} \int d^3\mathbf x \, T^{k_{n-1} k_n} r^2 x^{N-2} \, \partial_{N-2} E_{k_{n-1} k_n} \, .
\end{align}
The subscripts on the terms of the action denote the number of pairs of indices which can be antisymmetrized without necessarily vanishing in the action.
We see that $S_0$ is already symmetrized.

The strategy to write the action in terms of symmetric tensors is the following: First, we perform the (partial) symmetrization of $S_2$ reducing it to terms which have at most 1 pair of indices which may give a non-vanishing contribution to the action when antisymmetrized. Then we perform the symmetrization of all terms where one pair of indices can be antisymmetrized similarly to what we did for the electromagnetic case.

For the reduction of of $S_2$ we use the decomposition via Young tableaux from Eqs. (\ref{eq_YoungGR}) - (\ref{eq_YT3}) which, aside from the last line in Eq. (\ref{eq_YT1}) where we used the moment relations following from $\partial_\mu T^{\mu \nu} = 0$, hold in the same form when additional factors of $r^2$ are inserted in the integrals. Otherwise, the procedure works analogous to the electromagnetic case where we decomposed Eq. (\ref{eq_SB}) by using the decomposition with Young tableaux infinitely many times. In doing so, we also use the moment relation
\begin{align}
  \int d^3\mathbf x \left[T^{ab} r^{2p} x^N + \left(T^{ak_n} r^{2p} x^{bN-1} + \text{k-perms} \right)\right] & = \int d^3\mathbf x \dot T^{0a} r^{2p} x^{bN} - 2 p \int d^3\mathbf x T^{ac} r^{2(p-1)} x^{bcN} 
\end{align}
to rewrite the second Young tableau's expression in a more convenient form. After some quite lengthy manipulations, we obtain
\begin{align}
S_2 & =  \int dt \sum_{n=2}^{\infty} \sum_{q=0}^{\infty} \frac{n-1}{(n+2q+1)!} \left[\partial_t^{2q} \int d^3\mathbf x T^{k_{n-1} k_n} r^{2q+2} x^{N-2} \right]^\text{S} \partial_{N-2} E_{k_{n-1} k_n} \notag \\
& +  \int dt \sum_{n=2}^{\infty} \sum_{q=1}^{\infty} \frac{n-1}{(n+2q+1)!} \left[\partial_t^{2q} \int d^3\mathbf x T^{aa} r^{2q} x^{N} \right]^\text{S} \partial_{N-2} E_{k_{n-1} k_n} \notag \\
& -  \int dt \sum_{n=2}^{\infty} \sum_{q=1}^{\infty} \frac{2(n-1)}{(n+2q+1)!} \left\{\partial_t^{2q} \int d^3\mathbf x T^{ak_n} r^{2q} x^{aN-1} \right\} \partial_{N-2} E_{k_{n-1} k_n} \notag \\
& -  \int dt \sum_{n=2}^{\infty} \sum_{q=1}^{\infty} \frac{2n}{(n+2q)! (n+1)} \notag \\
& {} \hspace*{100pt}\int d^3\mathbf x \epsilon^{k_nba} \left\{ \partial_t^{2q} T^{0a} r^{2q} x^{bN-1} - 2q \partial_t^{2q-1} T^{ac} r^{2q-2} x^{bcN-1} \right\} \partial_{N-2} B_{k_{n-1} k_n} \, .
\end{align}
Plugging this into $S_{\text{source}}^{rad}$ we have
\begin{align}
S_{\text{source}}^{rad} & = \int dt \sum_{n=2}^{\infty} \frac{1}{n!} \left[\int d^3\mathbf x T^{00} x^{N} \right]^\text{S} \partial_{N-2} E_{k_{n-1} k_n} \notag \\
& +  \int dt \sum_{n=2}^{\infty} \sum_{q=0}^{\infty} \frac{n-1}{(n+2q+1)!} \left[\partial_t^{2q} \int d^3\mathbf x T^{k_{n-1} k_n} r^{2q+2} x^{N-2} \right]^\text{S} \partial_{N-2} E_{k_{n-1} k_n} \notag \\
& +  \int dt \sum_{n=2}^{\infty} \sum_{q=0}^{\infty} \frac{n-1}{(n+2q+1)!} \left[\partial_t^{2q} \int d^3\mathbf x T^{aa} r^{2q} x^{N} \right]^\text{S} \partial_{N-2} E_{k_{n-1} k_n} \notag \\
& -  \int dt \sum_{n=2}^{\infty} \sum_{q=0}^{\infty} \frac{2(n-1)}{(n+2q+1)!} \left\{\partial_t^{2q} \int d^3\mathbf x T^{ak_n} r^{2q} x^{aN-1} \right\} \partial_{N-2} E_{k_{n-1} k_n} \notag \\
& -  \int dt \sum_{n=2}^{\infty} \sum_{q=0}^{\infty} \frac{2n}{(n+2q)! (n+1)} \notag \\
& {} \hspace*{100pt} \int d^3\mathbf x \epsilon^{k_nba} \left\{ \partial_t^{2q} T^{0a} r^{2q} x^{bN-1} - 2q \partial_t^{2q-1} T^{ac} r^{2q-2} x^{bcN-1} \right\} \partial_{N-2} B_{k_{n-1} k_n} \label{eq_SGR_S2_sym}
\end{align}
where in the sums over $q$ in the last three terms the $q=0$ components were provided from $S_0$, $S_1^{E, \, 0}$ and $S_1^{B, \, 0}$ respectively. Only the last two expressions in Eq. (\ref{eq_SGR_S2_sym}) need to be further symmetrized, where the procedure works analogously to the E\&M calculation. For the fourth line of Eq. (\ref{eq_SGR_S2_sym}) the symmetrization procedure yields
\begin{align}
S_1^E & =  -  \int dt \sum_{n=2}^{\infty} \sum_{q=0}^{\infty} \frac{2(n-1)}{(n+2q+1)!} \left\{\int d^3\mathbf x \, \partial_t^{2q} T^{ak_n} r^{2q} x^{aN-1} \right\} \partial_{N-2} E_{k_{n-1} k_n} \notag \\
& = \int dt \sum_{n=2}^{\infty} \sum_{q=0}^{\infty}  \sum_{s=0}^{\infty} \frac{2 n (n-1)}{(n+2q+2s+3)!(n+2s+2)}  \notag \\
 & {} \hspace*{95pt} \left[ \int d^3\mathbf x \, \partial_t^{2q+2s+2} T^{ab} r^{2q+2s} x^{ab N}\right]^\text{S} \partial_{N-2} E_{k_{n-1} k_n} \notag \\
& - \int dt \sum_{n=2}^{\infty} \sum_{q=0}^{\infty} \sum_{s=0}^{\infty} \frac{2 n (n-1)}{(n+2q+2s+1)!(n+2s)} \notag \\
 & {} \hspace*{95pt} \left[ \int d^3\mathbf x \, \partial_t^{2q+2s} T^{ak_n} r^{2q+2s} x^{a N-1}\right]^\text{S} \partial_{N-2} E_{k_{n-1} k_n}   \notag \\
& + \int dt \sum_{n=2}^{\infty} \sum_{q=0}^{\infty} \sum_{s=0}^{\infty} \frac{2 n (n-1)}{(n+2q+2s+2)!(n+2s+1)} \notag \\
 & {} \hspace*{95pt} \left[ \int d^3\mathbf x \, \partial_t^{2q+2s+1} \epsilon^{k_nba}T^{ac} r^{2q+2s} x^{bc N-1}\right]^\text{S} \partial_{N-2} B_{k_{n-1} k_n} 
\end{align}
and for the last two lines of Eq. (\ref{eq_SGR_S2_sym}) it gives
\begin{align}
S_1^B & =  -  \int dt \sum_{n=2}^{\infty} \sum_{q=0}^{\infty} \frac{2n}{(n+2q)! (n+1)}  \notag \\
 & {} \hspace*{100pt}\int d^3\mathbf x \epsilon^{k_nba} \left\{ \partial_t^{2q} T^{0a} r^{2q} x^{bN-1} - 2q \partial_t^{2q-1} T^{ac} r^{2q-2} x^{bcN-1} \right\} \partial_{N-2} B_{k_{n-1} k_n} \notag \\
 & = - \int dt \sum_{n=2}^{\infty} \sum_{q=0}^{\infty} \sum_{s=0}^{\infty} \frac{2 n (n-1)}{(n+2q+2s+1)!(n+2s+2)(n+2s)} \notag \\
 & {} \hspace*{34pt} \left[ \int d^3\mathbf x \left( \partial_t^{2q+2s+1} T^{0a} r^{2q+2s} x^{a N} - 2q \partial_t^{2q+2s} T^{ab} r^{2q+2s-2} x^{abN}\right) \right]^\text{S} \partial_{N-2} E_{k_{n-1} k_n}  \notag \\
& + \int dt \sum_{n=2}^{\infty} \sum_{q=0}^{\infty} \sum_{s=0}^{\infty} \frac{2 n (n-1)}{(n+2q+2s+1)!(n+2s+2)(n+2s)} \notag \\
 & {} \hspace*{34pt} \left[ \int d^3\mathbf x \left( \partial_t^{2q+2s+1} T^{0k_n} r^{2q+2s+2} x^{N-1} - 2q \partial_t^{2q+2s} T^{ak_n} r^{2q+2s} x^{aN-1} \right) \right]^\text{S} \partial_{N-2} E_{k_{n-1} k_n}  \notag \\
& - \int dt \sum_{n=2}^{\infty} \sum_{q=0}^{\infty} \sum_{s=0}^{\infty} \frac{2 n (n-1)}{(n+2q+2s)!(n+2s+1)(n+2s-1)} \notag \\
 & {} \hspace*{34pt} \left[ \int d^3\mathbf x \epsilon^{k_nba} \left(\partial_t^{2q+2s} T^{0a} r^{2q+2s} x^{bN-1} - 2q \partial_t^{2q+2s-1} T^{ac} r^{2q+2s-2} x^{bcN-1} \right) \right]^\text{S} \partial_{N-2} B_{k_{n-1} k_n}  
\end{align}
so that we obtain an expression for $S_{\text{source}}^{rad}$ entirely in terms of symmetric tensors. Its form can be simplified by removing the two infinite sums over $q$ and $s$ in $S_1^E$ and $S_1^B$ using variable redefinitions $q \rightarrow q + s$ which yield one infinite sum over $q$ and a finite sum over $s$. The latter can all be performed using Eqs. (\ref{eq_Sums1}) and (\ref{eq_Sums2}) and we find
the final result for $S_{\text{source}}^{rad}$ in terms of symmetric tensors,
\begin{align}
S_{\text{source}}^{rad} & = \int dt \sum_{n=2}^{\infty} \frac{1}{n!} \left[\int d^3\mathbf x \, T^{00} x^{N} \right]^\text{S} \partial_{N-2} E_{k_{n-1} k_n} \notag \\
& +  \int dt \sum_{n=2}^{\infty} \sum_{q=0}^{\infty} \frac{n-1}{(n+2q+1)!} \left[\int d^3\mathbf x \, \partial_t^{2q} T^{aa} r^{2q} x^{N} \right]^\text{S} \partial_{N-2} E_{k_{n-1} k_n} \notag \\
& + \int dt \sum_{n=2}^{\infty} \sum_{q=0}^{\infty}  \frac{2 (n-1)(q+1)}{(n+2q+3)!} \left[ \int d^3\mathbf x \, \partial_t^{2q+2} T^{ab} r^{2q} x^{ab N}\right]^\text{S} \partial_{N-2} E_{k_{n-1} k_n} \notag \\
& - \int dt \sum_{n=2}^{\infty} \sum_{q=0}^{\infty}  \frac{2 (n-1)(q+1)}{(n+2q+2)!} \left[ \int d^3\mathbf x \, \partial_t^{2q+1} T^{0a} r^{2q} x^{a N} \right]^\text{S} \partial_{N-2} E_{k_{n-1} k_n}  \notag \\
& - \int dt \sum_{n=2}^{\infty} \sum_{q=0}^{\infty} \frac{2 (n-1)(q+1)}{(n+2q+1)!}  \left[ \int d^3\mathbf x \, \partial_t^{2q} T^{ak_n} r^{2q} x^{a N-1}\right]^\text{S} \partial_{N-2} E_{k_{n-1} k_n}   \notag \\
& + \int dt \sum_{n=2}^{\infty} \sum_{q=0}^{\infty} \frac{2 (n-1)(q+1)}{(n+2q+2)!} \left[ \int d^3\mathbf x \, \partial_t^{2q+1} T^{0k_n} r^{2q+2} x^{N-1} \right]^\text{S} \partial_{N-2} E_{k_{n-1} k_n}  \notag \\
& +  \int dt \sum_{n=2}^{\infty} \sum_{q=0}^{\infty} \frac{n-1}{(n+2q+1)!} \left[ \int d^3\mathbf x \, \partial_t^{2q} T^{k_{n-1} k_n} r^{2q+2} x^{N-2} \right]^\text{S} \partial_{N-2} E_{k_{n-1} k_n} \notag \\
& - \int dt \sum_{n=2}^{\infty} \sum_{q=0}^{\infty} \frac{2 n (q+1)}{(n+2q+1)!} \left[ \int d^3\mathbf x \, \epsilon^{k_nba} \partial_t^{2q} T^{0a} r^{2q} x^{bN-1} \right]^\text{S} \partial_{N-2} B_{k_{n-1} k_n} \notag \\
& + \int dt \sum_{n=2}^{\infty} \sum_{q=0}^{\infty} \frac{2 n (q+1)}{(n+2q+2)!}  \left[ \int d^3\mathbf x \, \epsilon^{k_nba} \partial_t^{2q+1} T^{ac} r^{2q} x^{bc N-1}\right]^\text{S} \partial_{N-2} B_{k_{n-1} k_n} \label{eq_SGR_S2_sym4}
\end{align}

The final step in the multipole expansion of the action is to take out the traces so that all coefficients are in terms of STF tensors. For this purpose, let us define 4 separate components of Eq. (\ref{eq_SGR_S2_sym4}). The first piece $S_1$ comprises the first four lines in Eq. (\ref{eq_SGR_S2_sym4}), the second piece $S_2$ are lines five and six, the third piece $S_3$ is line seven and the fourth piece $S_4$ is given by lines eight and nine. Each of these four pieces is written in terms of STF tensors using Eq. (\ref{MNSTFnew}), but the counting of resulting trace structures differs.

For the first piece $S_1$ all contracted indices in the STF prescription will give powers of $r^2$, so that the only counting left to do is to omit terms which vanish by $E_{ii} = \partial_i E_{ij} = 0$. This is similar to what we did for the electromagnetic case (where the $\nabla \cdot \mathbf E = 0$ components had to be omitted, see Eq. (\ref{eq_SE}) and the discussion above). Here this is accounted for by inserting a factor of
\begin{equation}
\frac{{n \choose 2p} - {n-1 \choose 2p-1} - {n-2 \choose 2p-1}}{{n \choose 2p}} \label{eq_countEiidiEij}
\end{equation}
into the STF decomposition of Eq. (\ref{MNSTFnew}) and only keeping the traces which don't yield $E_{ii}$ or $\partial_i E_{ij}$.
We obtain
\begin{align}
S_1 & = \int dt \sum_{\ell=2}^{\infty} \sum_{p=0}^{\infty} \frac{c_p^{(\ell+2p)}}{(\ell+2p)!} \frac{\ell (\ell-1)}{(\ell+2p)(\ell+2p-1)} \left[ \int d^3\mathbf x \, \partial_t^{2p}T^{00} r^{2p} x^{L} \right]^\text{STF} \partial_{L-2} E_{k_{\ell-1} k_\ell} \notag \\
& +  \int dt \sum_{\ell=2}^{\infty} \sum_{p=0}^{\infty} \frac{\ell(\ell-1)}{(\ell+2p+1)!} \left(\sum_{j=0}^{p} \frac{c_j^{(\ell+2j)}}{\ell+2j}\right) \left[\int d^3\mathbf x \, \partial_t^{2p} T^{aa} r^{2p} x^{L} \right]^\text{STF} \partial_{L-2} E_{k_{\ell-1} k_\ell} \notag \\
& +  \hspace*{-1pt}  \int \! dt \hspace*{-1pt} \sum_{\ell=2}^{\infty} \sum_{p=0}^{\infty}  \hspace*{-1pt}  \frac{2 \ell (\ell-1)}{(\ell \hspace*{-1pt} + \hspace*{-1pt} 2p \hspace*{-1pt} + \hspace*{-1pt} 3)!} \! \left(\sum_{j=0}^{p} \frac{(p \hspace*{-1pt} - \hspace*{-1pt} j \hspace*{-1pt} + \hspace*{-1pt} 1) c_j^{(\ell+2j)}}{\ell+2j}\right)  \hspace*{-1pt}  \!  \hspace*{-1pt} \left[ \int \! d^3\mathbf x \, \partial_t^{2p+2} T^{ab} r^{2p} x^{ab L} \hspace*{-1pt} \right]^\text{STF} \!  \hspace*{-1pt} \partial_{L-2} E_{k_{\ell-1} k_\ell} \notag \\
& -  \hspace*{-1pt} \int \! dt  \hspace*{-1pt} \sum_{\ell=2}^{\infty} \sum_{p=0}^{\infty}  \hspace*{-1pt}   \frac{2 \ell (\ell-1)}{(\ell \hspace*{-1pt} + \hspace*{-1pt} 2p \hspace*{-1pt} + \hspace*{-1pt} 2)!}  \! \left(\sum_{j=0}^{p}  \hspace*{-1pt} \frac{(p \hspace*{-1pt} - \hspace*{-1pt} j \hspace*{-1pt} + \hspace*{-1pt} 1) c_j^{(\ell+2j)}}{\ell+2j}\right) \! \! \left[ \int \! d^3\mathbf x \, \partial_t^{2p+1} T^{0a} r^{2p} x^{a L} \right]^\text{STF} \! \partial_{L-2} E_{k_{\ell-1} k_\ell} 
\end{align}
For the remaining pieces $S_2$, $S_3$ and $S_4$, we also need to count which of the different possible structures of traces in the moments occur how many times. For example, in the $T^{0k_n}$ part of $S_2$, the traces when applying the STF decomposition of Eq. (\ref{MNSTFnew}) can yield either $r^2$ or $T^{0a} \mathbf x^a$, and we need to know how many times each of them occurs. It is a straightforward counting exercise, and for the parts of the action $S_2$ and $S_4$ which have only one of the free indices $k_1 \dots k_n$ not on a vector $\mathbf x$, the counting is the same as in Eq. (\ref{eq_SB3}). Moreover, we note that the factor in Eq. (\ref{eq_countEiidiEij}) is universal for all terms and is to be used multiplicatively with the factors which account for the different traced structures in the moments. We obtain for the second part of the action
\begin{align}
S_2 & = - \int dt \sum_{\ell=2}^{\infty} \sum_{p=0}^{\infty}  \frac{4 \ell (\ell-1)}{(\ell+2p+1)!} \left(\sum_{j=0}^{p} \frac{j(p-j+1) c_j^{(\ell+2j)}}{(\ell+2j)^2}\right) \notag \\ 
& {} \hspace*{93pt}  \left[ \int d^3\mathbf x \, \partial_t^{2p} T^{ab} r^{2p-2} x^{ab L}\right]^\text{STF} \partial_{L-2} E_{k_{\ell-1} k_\ell} \notag \\
  & - \int dt \sum_{\ell=2}^{\infty} \sum_{p=0}^{\infty}  \frac{2 \ell^2 (\ell-1)}{(\ell+2p+1)!} \left(\sum_{j=0}^{p} \frac{(p-j+1) c_j^{(\ell+2j)}}{(\ell+2j)^2}\right)\notag \\ 
& {} \hspace*{93pt}   \left[ \int d^3\mathbf x \, \partial_t^{2p} T^{ak_\ell} r^{2p} x^{a L-1}\right]^\text{STF} \partial_{L-2} E_{k_{\ell-1} k_\ell} \notag \\
  & + \int dt \sum_{\ell=2}^{\infty} \sum_{p=0}^{\infty}  \frac{4 \ell (\ell-1)}{(\ell+2p+2)!} \left(\sum_{j=0}^{p} \frac{j(p-j+1) c_j^{(\ell+2j)}}{(\ell+2j)^2}\right) \notag \\ 
& {} \hspace*{93pt}  \left[ \int d^3\mathbf x \, \partial_t^{2p+1} T^{0a} r^{2p} x^{a L}\right]^\text{STF} \partial_{L-2} E_{k_{\ell-1} k_\ell} \notag \\
  & + \int dt \sum_{\ell=2}^{\infty} \sum_{p=0}^{\infty}  \frac{2 \ell^2 (\ell-1)}{(\ell+2p+2)!} \left(\sum_{j=0}^{p} \frac{(p-j+1) c_j^{(\ell+2j)}}{(\ell+2j)^2}\right) \notag \\ 
& {} \hspace*{93pt}  \left[ \int d^3\mathbf x \, \partial_t^{2p+1} T^{0k_\ell} r^{2p+2} x^{L-1}\right]^\text{STF} \partial_{L-2} E_{k_{\ell-1} k_\ell} \, ,
\end{align}
the third one becomes
\begin{align}
S_3 & = \int dt \sum_{\ell=2}^{\infty} \sum_{p=0}^{\infty}  \frac{2 \ell (\ell-1)}{(\ell+2p+1)!} \left(\sum_{j=0}^{p} \frac{j c_j^{(\ell+2j)}}{(\ell+2j)^2(\ell+2j-1)}\right) \notag \\ 
& {} \hspace*{93pt}   \left[ \int d^3\mathbf x \, \partial_t^{2p} T^{aa} r^{2p} x^{L}\right]^\text{STF} \partial_{L-2} E_{k_{\ell-1} k_\ell} \notag \\
& + \int dt \sum_{\ell=2}^{\infty} \sum_{p=0}^{\infty}  \frac{4 \ell (\ell-1)}{(\ell+2p+1)!} \left(\sum_{j=0}^{p} \frac{j (j-1) c_j^{(\ell+2j)}}{(\ell+2j)^2(\ell+2j-1)}\right) \notag \\ 
& {} \hspace*{93pt}   \left[ \int d^3\mathbf x \, \partial_t^{2p} T^{ab} r^{2p-2} x^{abL}\right]^\text{STF} \partial_{L-2} E_{k_{\ell-1} k_\ell} \notag \\
& + \int dt \sum_{\ell=2}^{\infty} \sum_{p=0}^{\infty}  \frac{\ell^2 (\ell-1)^2}{(\ell+2p+1)!} \left(\sum_{j=0}^{p} \frac{c_j^{(\ell+2j)}}{(\ell+2j)^2(\ell+2j-1)}\right) \notag \\ 
& {} \hspace*{93pt}  \left[ \int d^3\mathbf x \, \partial_t^{2p} T^{k_{\ell-1} k_\ell} r^{2p+2} x^{L-2}\right]^\text{STF} \partial_{L-2} E_{k_{\ell-1} k_\ell} \notag \\
& + \int dt \sum_{\ell=2}^{\infty} \sum_{p=0}^{\infty}  \frac{4 \ell^2 (\ell-1)}{(\ell+2p+1)!} \left(\sum_{j=0}^{p} \frac{j c_j^{(\ell+2j)}}{(\ell+2j)^2(\ell+2j-1)}\right) \notag \\ 
& {} \hspace*{93pt}  \left[ \int d^3\mathbf x \, \partial_t^{2p} T^{a k_\ell} r^{2p} x^{aL-1}\right]^\text{STF} \partial_{L-2} E_{k_{\ell-1} k_\ell} \, ,
\end{align}
and the fourth part reads
\begin{align}
S_4 & =  - \int dt \sum_{\ell=2}^{\infty} \sum_{p=0}^{\infty}  \frac{2 \ell^2 (\ell-1)}{(\ell+2p+1)!} \left(\sum_{j=0}^{p} \frac{(p-j+1) c_j^{(\ell+2j)}}{(\ell+2j)(\ell+2j-1)}\right) \notag \\ 
& {} \hspace*{93pt}  \left[ \int d^3\mathbf x \, \epsilon^{k_\ell ba} \partial_t^{2p} T^{0a} r^{2p} x^{bL-1}\right]^\text{STF} \partial_{L-2} B_{k_{\ell-1} k_\ell} \notag \\
& + \int dt \sum_{\ell=2}^{\infty} \sum_{p=0}^{\infty}  \frac{2 \ell^2 (\ell-1)}{(\ell+2p+2)!} \left(\sum_{j=0}^{p} \frac{(p-j+1) c_j^{(\ell+2j)}}{(\ell+2j)(\ell+2j-1)}\right) \notag \\ 
& {} \hspace*{93pt}  \left[ \int d^3\mathbf x \, \epsilon^{k_\ell ba} \partial_t^{2p+1} T^{ac} r^{2p} x^{bcL-1}\right]^\text{STF} \partial_{L-2} B_{k_{\ell-1} k_\ell} \, .
\end{align}
Adding these and simplifying we find for $S_{\text{source}}^{rad}$ in terms of STF tensors
\begin{align}
S_{\text{source}}^{rad} & = \int dt \sum_{\ell=2}^{\infty} \sum_{p=0}^{\infty} \frac{c_p^{(\ell+2p)}}{(\ell+2p)!} \frac{\ell (\ell-1)}{(\ell+2p)(\ell+2p-1)} \left[ \int d^3\mathbf x \, \partial_t^{2p}T^{00} r^{2p} x^{L} \right]^\text{STF} \partial_{L-2} E_{k_{\ell-1} k_\ell} \notag \\
& +  \int dt \sum_{\ell=2}^{\infty} \sum_{p=0}^{\infty} \frac{\ell(\ell-1)}{(\ell+2p+1)!} \left(\sum_{j=0}^{p} \frac{c_j^{(\ell+2j)}\left((\ell+2j)^2-\ell\right)}{(\ell+2j)^2(\ell+2j-1)}\right) \notag \\
 & {} \hspace*{90pt} \left[\int d^3\mathbf x \, \partial_t^{2p} T^{aa} r^{2p} x^{L} \right]^\text{STF} \partial_{L-2} E_{k_{\ell-1} k_\ell} \notag \\
& + \int dt \sum_{\ell=2}^{\infty} \sum_{p=0}^{\infty}  \frac{2 \ell (\ell-1)}{(\ell+2p+3)!} \left(\sum_{j=0}^{p+1} \frac{c_j^{(\ell+2j)} \left[\ell(\ell+2j-1)(p-j+1) - 2 j (\ell+j)\right] }{(\ell+2j)^2(\ell+2j-1)}\right) \notag \\
 & {} \hspace*{90pt}  \left[ \int d^3\mathbf x \, \partial_t^{2p+2} T^{ab} r^{2p} x^{ab L}\right]^\text{STF} \partial_{L-2} E_{k_{\ell-1} k_\ell} \notag \\
& - \int dt \sum_{\ell=2}^{\infty} \sum_{p=0}^{\infty}   \frac{2 \ell^2 (\ell-1)}{(\ell+2p+2)!} \left(\sum_{j=0}^{p} \frac{c_j^{(\ell+2j)} (p-j+1) }{(\ell+2j)^2}\right) \notag \\
 & {} \hspace*{90pt} \left[ \int d^3\mathbf x \, \partial_t^{2p+1} T^{0a} r^{2p} x^{a L} \right]^\text{STF} \partial_{L-2} E_{k_{\ell-1} k_\ell}  \notag \\
  & - \int dt \sum_{\ell=2}^{\infty} \sum_{p=0}^{\infty}  \frac{2 \ell^2 (\ell-1)}{(\ell+2p+1)!} \left(\sum_{j=0}^{p} \frac{c_j^{(\ell+2j)} \left[(\ell+2j-1)(p-j)+\ell-1\right] }{(\ell+2j)^2(\ell+2j-1)}\right) \notag \\
 & {} \hspace*{90pt}  \left[ \int d^3\mathbf x \, \partial_t^{2p} T^{ak_\ell} r^{2p} x^{a L-1}\right]^\text{STF} \partial_{L-2} E_{k_{\ell-1} k_\ell} \notag \\
  & + \int dt \sum_{\ell=2}^{\infty} \sum_{p=0}^{\infty}  \frac{2 \ell^2 (\ell-1)}{(\ell+2p+2)!} \left(\sum_{j=0}^{p} \frac{c_j^{(\ell+2j)} (p-j+1) }{(\ell+2j)^2}\right) \notag \\
 & {} \hspace*{90pt} \left[ \int d^3\mathbf x \, \partial_t^{2p+1} T^{0k_\ell} r^{2p+2} x^{L-1}\right]^\text{STF} \partial_{L-2} E_{k_{\ell-1} k_\ell} \notag \\
& + \int dt \sum_{\ell=2}^{\infty} \sum_{p=0}^{\infty}  \frac{\ell^2 (\ell-1)^2}{(\ell+2p+1)!} \left(\sum_{j=0}^{p} \frac{c_j^{(\ell+2j)}}{(\ell+2j)^2(\ell+2j-1)}\right) \notag \\
 & {} \hspace*{90pt} \left[ \int d^3\mathbf x \, \partial_t^{2p} T^{k_{\ell-1} k_\ell} r^{2p+2} x^{L-2}\right]^\text{STF} \partial_{L-2} E_{k_{\ell-1} k_\ell} \notag \\
& - \int dt \sum_{\ell=2}^{\infty} \sum_{p=0}^{\infty}  \frac{2 \ell^2 (\ell-1)}{(\ell+2p+1)!} \left(\sum_{j=0}^{p} \frac{(p-j+1) c_j^{(\ell+2j)}}{(\ell+2j)(\ell+2j-1)}\right) \notag \\
 & {} \hspace*{90pt} \left[ \int d^3\mathbf x \, \epsilon^{k_\ell ba} \partial_t^{2p} T^{0a} r^{2p} x^{bL-1}\right]^\text{STF} \partial_{L-2} B_{k_{\ell-1} k_\ell} \notag \\
& + \int dt \sum_{\ell=2}^{\infty} \sum_{p=0}^{\infty}  \frac{2 \ell^2 (\ell-1)}{(\ell+2p+2)!} \left(\sum_{j=0}^{p} \frac{(p-j+1) c_j^{(\ell+2j)}}{(\ell+2j)(\ell+2j-1)}\right) \notag \\
 & {} \hspace*{90pt} \left[ \int d^3\mathbf x \, \epsilon^{k_\ell ba} \partial_t^{2p+1} T^{ac} r^{2p} x^{bcL-1}\right]^\text{STF} \partial_{L-2} B_{k_{\ell-1} k_\ell} \, \label{eq_SgrSTF1}.
\end{align}
While the above Eq. (\ref{eq_SgrSTF1}) is in the desired form in terms of STF moments, it is still much too complicated. In order to simplify and to get rid of the sums over $j$, we first use the moment relations resulting from $\partial_\mu T^{\mu \nu} = 0$ to simplify the expressions for the electric moments. In particular, we first use 
\begin{align}
  \left[\int d^3\mathbf x \hspace*{1pt} T^{k_{\ell-1} k_\ell} r^{2p+2} x^{L-2}\right]^\text{S} & = \frac{1}{\ell-1} \left[\int d^3\mathbf x \hspace*{1pt} \dot T^{0k_\ell} r^{2p+2} x^{L-1}\right]^\text{S} - \frac{ 2 p +2}{\ell-1} \left[\int d^3\mathbf x \hspace*{1pt} T^{ak_\ell} r^{2p} x^{aL-1}\right]^\text{S}
\end{align}
in order to replace the components with $T^{k_{\ell-1} k_\ell}$, and subsequently we apply
\begin{align}
  \left[\int d^3\mathbf x \hspace*{1pt} T^{a k_\ell} r^{2p} x^{aL-1}\right]^\text{S} & = \frac{1}{\ell} \left[\int d^3\mathbf x \hspace*{1pt} \dot T^{0a} r^{2p} x^{aL}\right]^\text{S} - \frac{ 2 p}{\ell} \left[\int d^3\mathbf x \hspace*{1pt} T^{ab} r^{2p-2} x^{abL}\right]^\text{S} \notag \\
  & - \frac{1}{\ell} \left[\int d^3\mathbf x \hspace*{1pt} T^{aa} r^{2p} x^{L}\right]^\text{S} 
\end{align}
and 
\begin{align}
  \left[\int d^3\mathbf x \hspace*{1pt} T^{0 k_\ell} r^{2p+2} x^{L-1}\right]^\text{S} & = \frac{1}{\ell} \left[\int d^3\mathbf x \hspace*{1pt} \dot T^{00} r^{2p+2} x^{L}\right]^\text{S} - \frac{ 2 p + 2}{\ell} \left[\int d^3\mathbf x \hspace*{1pt} T^{0a} r^{2p} x^{aL}\right]^\text{S} 
\end{align}
to replace all structures with $T^{a k_\ell}$ and $T^{0 k_\ell}$. 

After these replacements, we can perform all remaining sums over $j$ using Eqs. (\ref{eq_Sum1}) - (\ref{eq_Sum3}), and arrive at our final result for the multipole expanded source action
\begin{align}
 S_{\text{source}} & = - \frac{1}{2 m_{Pl}} \int dt \left[M  h_{00} + 2 \mathbf P^i h_{0i} + M \mathbf X^i \partial_i h_{00} + \mathbf L^i \epsilon_{ijk} \partial_j h_{0k}\right] \notag \\
   & + \int dt \sum_{\ell = 2}^\infty \frac{1}{\ell!} \hspace*{1.5pt} I^L \, \partial_{L-2} E_{k_{\ell-1} k_\ell} - \int dt \sum_{\ell = 2}^\infty \frac{2 \ell}{(\ell+1)!} \hspace*{1.5pt} J^L \, \partial_{L-2} B_{k_{\ell-1} k_\ell} \label{eq_SmpxGRlin}
\end{align}
with the exact expressions for the multipole moments
\begin{align}
M & = \int d^3 \mathbf x \, T^{00} \label{eq_resultM}\\
\mathbf P^i & = \int d^3 \mathbf x \, T^{0i} \\
M \mathbf X^i & = \int d^3 \mathbf x \, T^{00} \mathbf x^i\\
\mathbf L^i & = - \int d^3 \mathbf x \, \epsilon^{ijk} T^{0j} \mathbf x^k \\
I^L & = \sum_{p=0}^\infty \frac{(2 \ell + 1)!!}{(2 p)!! (2 \ell + 2p + 1)!!} \left(1 + \frac{8 p \hspace*{1pt} (\ell + p + 1)}{(\ell + 1)(\ell + 2)}\right) \left[\int d^3  \mathbf x \, \partial_t^{2p} T^{00} r^{2 p} x^L\right]^{\text{STF}} \notag \\
& + \sum_{p=0}^\infty \frac{(2 \ell + 1)!!}{(2 p)!! (2 \ell + 2p + 1)!!} \left(1 + \frac{4 p}{(\ell + 1)(\ell + 2)}\right) \left[\int d^3  \mathbf x \, \partial_t^{2p} T^{aa} r^{2 p} x^L\right]^{\text{STF}} \notag \\
& - \sum_{p=0}^\infty \frac{(2 \ell + 1)!!}{(2 p)!! (2 \ell + 2p + 1)!!} \, \frac{4}{\ell + 1} \left(1 + \frac{2 p}{\ell + 2}\right) \left[\int d^3  \mathbf x \, \partial_t^{2p+1} T^{0a} r^{2 p} x^{aL}\right]^{\text{STF}} \notag \\
& + \sum_{p=0}^\infty \frac{(2 \ell + 1)!!}{(2 p)!! (2 \ell + 2p + 1)!!} \, \frac{2}{(\ell + 1)(\ell + 2)} \left[\int d^3  \mathbf x \, \partial_t^{2p+2} T^{ab} r^{2 p} x^{abL}\right]^{\text{STF}} \label{eq_resultIL} \\
J^L & = \sum_{p=0}^\infty \frac{(2 \ell + 1)!!}{(2 p)!! (2 \ell + 2p + 1)!!} \left(1 + \frac{2 p}{\ell + 2}\right) \left[ \int d^3\mathbf x \, \epsilon^{k_\ell ba} \partial_t^{2p} T^{0a} r^{2p} x^{bL-1}\right]^\text{STF}  \notag \\
 & - \sum_{p=0}^\infty \frac{(2 \ell + 1)!!}{(2 p)!! (2 \ell + 2p + 1)!!} \,\frac{1}{\ell + 2} \left[ \int d^3\mathbf x \, \epsilon^{k_\ell ba} \partial_t^{2p+1} T^{ac} r^{2p} x^{bcL-1}\right]^\text{STF}  \, . \label{eq_resultJL}
\end{align}
The above form of the multipole moments $I^L$ and $J^L$ is not unique and can be modified using moment relations derived from $\partial_\mu T^{\mu \nu} = 0$.
See the appendix \ref{app_powercntng} for the power counting rules which tell us which terms are required at a given order in PN calculations using NRGR.

\subsection{Energy flux in gravitational waves}
In order to calculate the energy flux or radiated power, we follow the same procedure explained above in the scalar and E\&M sections. We here work in linearized gravity neglecting nonlinearities, see \cite{Goldberger:2009qd} for calculations in the EFT framework of the leading effect of nonlinearities in the gravitational wave energy flux. We use a physical gauge with polarization tensors satisfying $\epsilon^{0\mu}(\mathbf k, h) =  \epsilon^{ii}(\mathbf k, h) = \mathbf k^i \epsilon^{ij}(\mathbf k, h) = 0$ and where the helicity sum is \cite{Goldberger:2009qd} 
\begin{align}
\sum_h \epsilon^{ij}({\mathbf k},h) \epsilon^{kl*}({\bf k}, h)&=  \frac{1}{2} \left[\delta^{ik}\delta^{jl}+\delta^{il}\delta^{jk}-\delta^{ij}\delta^{kl} +\frac{1}{\mathbf k^2} \left(\delta^{ij} \mathbf k^k \mathbf k^l +\delta^{kl} \mathbf k^i \mathbf k^j\right)\right. \notag \\
& \left. {} \hspace*{22pt} -\frac{1}{\mathbf k^2} \left(\delta^{ik} \mathbf k^j \mathbf k^l + \delta^{il} \mathbf k^j \mathbf k^k + \delta^{jk} \mathbf k^i \mathbf k^l + \delta^{jl} \mathbf k^i \mathbf k^k\right) + \frac{1}{\mathbf k^4} \mathbf k^i \mathbf k^j \mathbf k^k \mathbf k^l \right] \label{eq_GRpolSum} \, .
\end{align}
In this gauge there is only $h_{ij}$ emission, and the amputated on-shell graviton emission amplitude is
\begin{align}
 i \mathcal A_h & = \frac{i}{2 m_{Pl}} \sum_{\ell = 2}^\infty  \frac{(-i)^{\ell-2}}{\ell!} I^L(|\mathbf k|) \, k^{L-2} |\mathbf k|^2 \epsilon^{k_{\ell-1} k_{\ell} *}(\mathbf k, h) \notag \\
  & - \frac{i}{2 m_{Pl}} \sum_{\ell = 2}^\infty (-i)^{\ell -2} \frac{2  \ell}{(\ell + 1)!}  J^L(|\mathbf k|) \, k^{jL-2} |\mathbf k| \epsilon^{i j k_{\ell - 1}} \epsilon^{k_\ell i *}(\mathbf k, h) \, .
\end{align}

Upon squaring of the amplitude, summing over helicities using Eq. (\ref{eq_GRpolSum}) and evaluating the angular integrals with Eq. (\ref{eq_AngInt}), we find the energy flux
\begin{align}
 \dot E & =  \frac{G}{\pi T} \int_0^\infty dk \left[\sum_{\ell=2}^{\infty} \frac{(\ell+1)(\ell+2)}{\ell (\ell - 1) \hspace*{1pt} \ell! (2 \ell + 1)!!} \  k^{2 (\ell+1)} \left| I^L (k)\right|^2  \right. \notag \\
 & {} \hspace*{65pt} + \left. \sum_{\ell=2}^{\infty}   \frac{4 \ell (\ell + 2)}{(\ell - 1) \hspace*{1pt} (\ell+1)! (2 \ell + 1)!!} \  k^{2 (\ell+1)} \left| J^L (k)\right|^2 \right] \notag \\
            & =  \sum_{\ell=2}^{\infty} \frac{G (\ell + 1)(\ell + 2)} {\ell (\ell - 1) \hspace*{1pt} \ell! \hspace*{1pt} (2 \ell + 1)!!} \left< \left(\frac{d^{\ell+1}}{dt^{\ell+1}} I^L\right)^2\right>  + \sum_{\ell=0}^{\infty}  \frac{4 G \hspace*{1pt} \ell (\ell + 2)}{(\ell - 1) \hspace*{1pt} (\ell + 1)! \hspace*{1pt} (2 \ell + 1)!!} \left< \left(\frac{d^{\ell+1}}{dt^{\ell+1}} I^L\right)^2\right> \label{eq_powerGR} 
\end{align}
which is of course gauge invariant and reproduces the well-known result of \cite{Thorne:1980ru}.

\subsection{Gravitational waveform}
The gravitational radiation field far away from the source is computed in harmonic gauge where the retarded propagator in $d=4$ spacetime dimensions is
\begin{equation}
 \frac{- i P_{\mu \nu \alpha \beta} \hspace*{1pt} \theta(t_f-t_i) }{4 \pi |\mathbf x_f - \mathbf x_i|} \hspace*{1pt} \delta(t_f - t_i - |\mathbf x_f - \mathbf x_i|) 
\end{equation}
with $P_{\mu \nu \alpha \beta} = \frac{1}{2} \left(\eta_{\mu \alpha} \eta_{\nu \beta} + \eta_{\mu \beta} \eta_{\nu \alpha} - \eta_{\mu \nu} \eta_{\alpha \beta}\right)$. The physical waveform is transverse traceless and is obtained from the harmonic gauge radiation field via contraction with the transverse-traceless projection operator. Conforming to the usual dimensionless normalization, we define the transverse traceless waveform to be
\begin{equation}
 h_{ij}^{TT}(x) \equiv  \frac{1}{m_{Pl}} \Lambda_{ij,kl} \, h_{kl}(x)
\end{equation}
with the transverse traceless projection operator given by 
\begin{equation}
 \Lambda_{ij,kl} (x) = \left(\delta_{ik} - \mathbf n_i \mathbf n_k\right) \left(\delta_{jl} - \mathbf n_j \mathbf n_l\right) - \frac{1}{2} \left(\delta_{ij} - \mathbf n_i \mathbf n_j\right) \left(\delta_{kl} - \mathbf n_k \mathbf n_l\right) \, .
 \end{equation}
We find the waveform to all orders in the multipole expansion in linearized gravity to be 
\begin{align}
 h_{ij}^{TT}(x) & = - \frac{4 G}{|\mathbf x|} \Lambda_{ij,k_{\ell-1} k_\ell} \! \left[ \sum_{\ell = 2}^{\infty} \frac{1}{\ell!} \, \partial_t^{\ell} I^{L} (t_{\text{ret}}) \, \mathbf n^{L-2} - \! \sum_{\ell = 2}^{\infty} \frac{2 \ell}{(\ell+1)!} \,\epsilon^{ab(k_\ell}  \partial_t^{\ell} J^{k_{\ell-1})bL-2} (t_{\text{ret}}) \, \mathbf n^{aL-2} \right] \, , \label{eq_WF}
\end{align}
where we point out that our overall sign differs from the (otherwise equivalent) expressions for $h_{ij}^{TT}$ in \cite{Blanchet:2002av} because we use a different metric signature.
Moreover, we can also derive the energy flux of Eq. (\ref{eq_powerGR}) from  the waveform in Eq. (\ref{eq_WF}) using 
\begin{equation}
 \dot E = \frac{1}{32 \pi G} \int d\Omega \, |\mathbf x|^2 \left<\dot h_{ij}^{TT} \dot h_{ij}^{TT}\right> \, .
\end{equation}

\subsection{Going beyond a linearized source term} \label{sec_GRbeyond}
Our result for the multipole expanded action for linearized gravity in Eq. (\ref{eq_SmpxGRlin}) with the moments in Eqs. (\ref{eq_resultM} - \ref{eq_resultJL}) can simply be extended beyond the linear regime using the diffeomorphism invariance of the action. In NRGR this is guaranteed to hold in the action of the effective radiation theory by the use of the background field method when integrating out the potential modes \cite{Goldberger:2004jt}. 

Operationally, covariantizing the action corresponds to replacing all Riemann tensors with their expressions to all orders in the gravitational field $h_{\mu \nu}$ and replacing all partial derivatives by covariant derivatives in Eq. (\ref{eq_SmpxGRlin}), as well as extending the couplings in $S_{\text{source}}^{cons}$ beyond the linear regime. We then have the form of the action written down based on symmetry arguments in \cite{Goldberger:2009qd}, which in the center of mass frame is
\begin{align}
 S_{\text{source}} & = - M \int d\tau - \frac{1}{2} \int dx^\mu \hspace*{1pt} \mathbf L^{i} \hspace*{1pt} \epsilon_{ijk} \omega_\mu^{jk} \notag \\
   & + \int d\tau \sum_{\ell = 2}^\infty \frac{1}{\ell!} \hspace*{1.5pt} I^L \, \nabla_{L-2} E_{k_{\ell-1} k_\ell} - \int d\tau \sum_{\ell = 2}^\infty \frac{2 \ell}{(\ell+1)!} \hspace*{1.5pt} J^L \, \nabla_{L-2} B_{k_{\ell-1} k_\ell} \label{eq_SmpxGRNlin}
\end{align}
where we can also replace $d \tau = dt \sqrt{g_{00}}$ and $dx^\mu = dt \delta^{\mu 0}$ when we parameterize the worldline in terms of coordinate time $t$. The Wilson coefficients of this action with couplings beyond the linear regime in Eq. (\ref{eq_SmpxGRNlin}) are still given by Eqs. (\ref{eq_resultM} - \ref{eq_resultJL}), the same multipole moments as the ones in the linear multipole expanded action in Eq. (\ref{eq_SmpxGRlin}).

However, we note that new couplings quadratic in $h_{\mu \nu}$ can be written down which are not included in Eq. (\ref{eq_SmpxGRNlin}), for examples terms proportional to $E_{ij} E_{kl}$. Their coefficients are not fixed by our work. Moreover, we expect that there exist some relationships between the one-graviton-emission amplitudes used to match at the linear level and multi-graviton-emission amplitudes, since they must conspire to combine in such a way that they provide the non-linear pieces in Eq. (\ref{eq_SmpxGRNlin}).

\section{Conclusions}
We have performed the multipole expansion of a linear source term action in the long wavelength approximation for three theories of increasing complexity, namely a scalar field, electromagnetism and general relativity. Throughout, we worked at the level of the action which we expressed in terms of manifestly gauge invariant operators in electromagnetism and general relativity. We provided the exact results for the multipole moments, and we gave the emitted radiation power and the radiation field far away from the source to all orders in the multipole expansion. Our results for the multipole moments agree in the appropriate long wavelength limit with the results in \cite{Damour:1990gj}.

Our results for linearized\footnote{See the discussion in Sec. \ref{sec_gr} for what precisely we mean by linearized.} gravity are an important component of the effective field theory framework NRGR. Our exact expressions for the multipole moments in Eqs. (\ref{eq_resultM} - \ref{eq_resultJL}) together with the expressions for the power emitted in gravitational waves in Eq. (\ref{eq_powerGR}) and the waveform in Eq. (\ref{eq_WF}) will greatly simplify future calculation of precision gravitational wave observables in the effective field theory framework. From now on, all that is left to do in the matching to the effective radiation theory is to compute the Feynman diagrams which provide the moments of $T^{\mu \nu}$ in the expressions for the multipole moments to a desired order. See the appendix \ref{app_powercntng} for the power counting rules which tell us which terms are required at a given order.

While we have worked with a linear source action in gravity,  our results hold more generally for the covariantized form of our multipole expanded action in Eq. (\ref{eq_SmpxGRNlin}) by diffeomorphism invariance. However, at quadratic order in the gravitational field, new invariant operators can be written down, and this work does not fix their coefficeints. The multipole expansion to all orders in the nonlinearities has been solved at the level of the solution to the equations of motion in the more traditional approach to compute gravitational wave signatures by Blanchet in \cite{Blanchet:1998in}, where the introduction of four additional sets of multipole moments was required. It would be an interesting future direction to further investigate the structure of the source action and the multipole expansion beyond the linear regime and its relation to the results of \cite{Blanchet:1998in}.

\begin{center}
{\bf Acknowledgements}
\end{center} 
I would like to thank Jui-yu Chiu, Hael Collins, Walter Goldberger, Rich Holman, Ambar Jain, Duff Neill, Rafael Porto and Ira Rothstein for useful conversations, and an anonymous referee and Shahar Hadar for bringing some typos to my attention. This work was supported by NASA grant 22645.1.1110173.

\appendix

\section{Useful formulas for sums}
Recalling that
\begin{equation}
 c_j^{(\ell + 2j)} = \frac{(\ell + 2 j)!}{\ell !} \hspace*{1pt} \frac{(2\ell + 1)!!}{(2j)!! (2 \ell + 2 j + 1)!!} 
\end{equation}
as defined above in  of Eq. (\ref{eq_c}), we have the sums
\begin{align}
\sum_{j=0}^{p} \frac{c_j^{(\ell+2j)}}{\ell + 2 j} & = \frac{1}{\ell} \hspace*{1pt} \frac{(\ell+2 p +1)!}{(\ell+1)!} \hspace*{1pt}  \frac{(2 \ell + 1)!!}{(2 p)!! (2 \ell + 2 p + 1)!!} \label{eq_Sum1} \\
\sum_{j=0}^{p} \frac{(p-j+1) c_j^{(\ell+2j)}}{(\ell + 2 j)(\ell+2j-1)} & = \frac{1}{\ell(\ell-1)} \hspace*{1pt} \frac{(\ell+2 p +2)!}{(\ell+2)!} \hspace*{1pt}  \frac{(2 \ell + 1)!!}{(2 p)!! (2 \ell + 2 p + 1)!!} \label{eq_Sum2} \\
\sum_{j=0}^{p} \frac{c_j^{(\ell+2j)}}{(\ell + 2 j)(\ell+2j-1)} & = \frac{1}{\ell(\ell-1)} \left[ \frac{(\ell+2 p)!}{\ell!} + 4 p  \frac{(\ell+2 p)!}{(\ell+2)!} \right]  \frac{(2 \ell + 1)!!}{(2 p)!! (2 \ell + 2 p + 1)!!} \label{eq_Sum3} \\
\sum_{s=0}^{q} \frac{1}{(n+2s+2)(n+2s)} & = \frac{q+1}{n(n+2q+2)} \label{eq_Sums1}\\
\sum_{s=0}^{q} \frac{1}{(n+2s+1)(n+2s-1)} & = \frac{q+1}{(n-1)(n+2q+1)} \label{eq_Sums2} \, .
\end{align}

\section{Power counting for radiation observables in NRGR} \label{app_powercntng}
The multipole moments in Eqs. (\ref{eq_resultM} - \ref{eq_resultJL}) together with the energy flux in Eq. (\ref{eq_powerGR}) and the waveform in Eq. (\ref{eq_WF}) greatly simplify future higher order calculation in the EFT framework NRGR. Here we briefly outline which multipole moments and need to be included for a calculation at a given order, and which of the terms in the infinite expressions for the multipole moments are required.

Generally, all time derivatives in the multipole moments or in the expressions for the power and the waveform scale as $v/a$ where $v$ is the typical three-velocity within the source. All factors of $\mathbf x$ in the moments scale as the size of the source $a$ (which is the orbital separation for binaries). Therefore, the next order in $p$ in the sums in the multipole moments in Eqs. (\ref{eq_resultIL}) and (\ref{eq_resultJL}), which always is accompanied by an extra factor of $\partial_t^2 r^2$, is suppressed by a factor $v^2$. The only ingredient we are missing in order to be able to power count everything are the moments. Their scaling is not universal and changes for example if we consider systems with spin. See Table \ref{tabl_Tscalings} \cite{Porto:2010zg} for their scalings where we defined $K^{\mu \nu}_\ell \equiv \int d^3x T^{\mu \nu} x^L$.

For practical purposes it is convenient to simply power count factors of $v$ with respect to the leading expression for a multipole moment or an observable. Let us discuss a quick example for a system where we can neglect spin: If we want for example the quadrupole moment to 2PN, i.e. at order $v^4$ beyond its leading expression, we need the $T^{00}$ component in Eq. (\ref{eq_resultIL}) to second order in $p$, the $T^{aa}$ and $T^{0a}$ components in Eq. (\ref{eq_resultIL}) to first order in $p$ and the $T^{ab}$ component in Eq. (\ref{eq_resultIL}) to zeroth order in $p$. Recall that each order in $p$ contributes to all orders in $v^2$ beyond its leading expression due to corrections from matching, see \cite{Goldberger:2009qd, Porto:2010zg} for examples and more details on matching.
  
{

\begin{center}
 \begin{table}
  \begin{tabular}{| r | c | c | c |}
    \hline
    {} & ${\cal O}(\not {\hspace*{-2pt}\mathbf S})$ & ${\cal O}({\bf S}_A)$ & ${\cal O}({\bf S}_A^2)$\\ \hline
    $K^{00}_\ell$ & $m a^\ell$ & $m a^\ell v^3$ & $m a^\ell v^4$ \\ \hline
    $K^{0i}_\ell$ & $m a^\ell v$ & $m a^\ell v^2$ & $m a^\ell v^5$ \\ \hline    
    $K^{ij}_\ell$ & $m a^\ell v^2$ & $m a^\ell v^3$ & $m a^\ell v^6$ \\ \hline    
  \end{tabular}
  \caption{Scalings for leading terms in moments of $T^{\mu \nu}$ for various orders in spin \cite{Porto:2010zg}. These scalings are valid for all moments $\ell \geq 2$ and persist if some indices are contracted.}
  \label{tabl_Tscalings}
 \end{table}
\end{center}

\end{document}